\documentclass[12pt]{article}

\usepackage{array}
\usepackage{multirow}
\usepackage{booktabs} 
\usepackage{bm}
\usepackage{mathtools}
\usepackage{booktabs}
\usepackage{amssymb}
\usepackage{accents}
\usepackage{amsthm}
\usepackage[english]{babel} 
\usepackage[protrusion=true,expansion=true]{microtype} 
\usepackage{microtype} 
\usepackage{tabularx}
\usepackage[font = small,labelfont=bf,textfont=it]{caption} 
\usepackage[title]{appendix}
\usepackage{dsfont}
\usepackage{booktabs} 
\usepackage{natbib}
\usepackage{setspace}
\usepackage{soul}
\usepackage{enumitem}

\usepackage{multirow}
\usepackage{algorithm}
\usepackage[noend]{algpseudocode}
\usepackage{etoolbox}
\algblock{ParFor}{EndParFor}
\algnewcommand\algorithmicparfor{\textbf{for}}
\algnewcommand\algorithmicpardo{}
\algnewcommand\algorithmicendparfor{}
\algrenewtext{ParFor}[1]{\algorithmicparfor\ #1\ }
\algrenewtext{EndParFor}{\algorithmicendparfor}
\newcolumntype{C}[1]{>{\centering\let\newline\\\arraybackslash\hspace{0pt}}m{#1}}

\usepackage{xcolor}

\newcommand{\RR}{\mathbb{R}}
\newcommand{\argmin}{\operatorname*{argmin}}
\newcommand{\argmax}{\operatorname*{argmax}}
\newcommand{\stb}{\State $\bullet$ \;}
\newcommand{\ubar}[1]{\underaccent{\bar}{#1}}

\usepackage{algorithm}
\usepackage{algpseudocode}

\usepackage{graphicx,epstopdf}
\makeatletter
\newcommand{\distas}[1]{\mathbin{\overset{#1}{\kern\z@\sim}}}%
\newsavebox{\mybox}\newsavebox{\mysim}
\newtheorem{theorem}{Theorem}[section]
\newtheorem{corollary}[theorem]{Corollary}
\newtheorem{proposition}[theorem]{Proposition}
\newtheorem{lemma}[theorem]{Lemma}

\newtheorem{remark}[theorem]{Remark}
\newcommand{\distras}[1]{%
  \savebox{\mybox}{\hbox{\kern3pt$\scriptstyle#1$\kern3pt}}%
  \savebox{\mysim}{\hbox{$\sim$}}%
  \mathbin{\overset{#1}{\kern\z@\resizebox{\wd\mybox}{\ht\mysim}{$\sim$}}}%
}
\newcolumntype{C}[1]{>{\centering\let\newline\\\arraybackslash\hspace{0pt}}m{#1}}

\newcommand{\blind}{1}

\begin{document}

\def\spacingset#1{\renewcommand{\baselinestretch}%
{#1}\small\normalsize} \spacingset{1.1}


\if1\blind
{
 
 \centering{\bf\Large Mesh-clustered Gaussian process emulator for partial differential equation boundary value problems}\\
  \vspace{0.2in}
  \centering{Chih-Li Sung$^{*,a}$, Wenjia Wang\footnote{These authors contributed equally to the manuscript. CS gratefully acknowledges funding from NSF DMS 2113407. WW gratefully acknowledges funding from NSFC grant 12101149.}$^{,b}$, Liang Ding$^{c}$, Xingjian Wang$^{d}$\vspace{0.2in}\\
        $^{a}$Michigan State University\\
    $^{b}$Hong Kong University of Science and Technology  (Guangzhou)\\
    $^{c}$Fudan University\quad
    $^{d}$Tsinghua University \\    }
    \date{\vspace{-7ex}}
} \fi

\if0\blind
{
  \bigskip
  \bigskip
  \bigskip
    \begin{center}
    {\Large\bf Mesh-clustered Gaussian process emulator for partial differential equation boundary value problems}
\end{center}
  \medskip
} \fi

\bigskip
\begin{abstract}

Partial differential equations (PDEs) have become an essential tool for modeling complex physical systems.  Such equations are typically solved numerically via mesh-based methods, such as finite element methods, with solutions over the spatial domain. However, obtaining these solutions are often prohibitively costly, limiting the feasibility of exploring parameters in PDEs. 
In this paper, we propose an efficient emulator that simultaneously predicts the solutions over the spatial domain, with theoretical justification of its uncertainty quantification. The novelty of the proposed method lies in the incorporation of the mesh node coordinates into the statistical model. In particular, the proposed method segments the mesh nodes into multiple clusters via a Dirichlet process prior and fits  Gaussian process models with the same hyperparameters in each of them. Most importantly, by revealing the underlying clustering structures, the proposed method can provide valuable insights into qualitative features of the resulting dynamics that can be used to guide further investigations. 
Real examples are demonstrated to show that our proposed method has smaller prediction errors than its main competitors, with competitive computation time, and identifies interesting clusters of mesh nodes that possess physical significance, such as satisfying boundary conditions.  An \textsf{R} package for the proposed methodology is provided in an open repository.
\end{abstract}

\noindent%
\vfill

\newpage
\spacingset{1.35} 

\section{Introduction}
Computer models have become essential to study physical systems that are expensive or infeasible, and have been successfully applied in a variety of scientific research, ranging from cell adhesion \citep{sung2020calibration} to designing a rocket injector \citep{mak2017efficient}. Typically, a physical system is described by a computer model consisting of a series of partial differential equations (PDEs) \citep{evans2010partial} and evaluated in a two- or three-dimensional space. For instance, in \cite{wang2018evolution}, the governing PDEs are evaluated to simulate the characteristics of a three-dimensional swirling flow in a cylindrical chamber. 

These PDEs are typically  solved   by numerical methods, such as finite element methods (FEMs) or a collocation method \citep{fornberg2015solving}, based on a mesh specification in a two- or three-dimensional space, and the numerical solutions are evaluated at these mesh node coordinates  (also called \textit{grid points} \citep{mak2017efficient,tan2018gaussian}).  The number of nodes is usually fairly large to ensure the numerical accuracy of  PDE solutions. Such evaluations, however, are often prohibitively costly for input space exploration. For instance, the high-fidelity simulation in \cite{mak2017efficient} generates around 100,000 grid points and takes six days of computation time for a given  input. Thus, it is essential to develop a cheaper statistical \textit{emulator} as a surrogate model that  approximates the solutions  in a timely fashion.



The paper focuses on developing an efficient emulator for a series of solutions at  \textit{many} coordinates in a \textit{PDE boundary value problem}, where the coordinates are fixed across inputs. A PDE boundary value problem involves solving a set of PDEs subject to specified conditions at the boundaries of the domain. A popular statistical emulator  for computer simulations is through Gaussian process (GP) modeling \citep{santner2013design,gramacy2020surrogates}, which provides a flexible approximation to the relationship between simulation output and inputs and quantifies uncertainty through its predictive variance; however, the GP is mainly for predicting a scalar output and is not directly applicable to the context of many outputs.  
One idea is to simultaneously emulate the output at each coordinate separately using independent GPs, which is discussed in \cite{qian2008gaussian}, \cite{conti2010bayesian}, and \cite{lee2011emulation, lee2012mapping}, and another idea is  using GPs with a special shared covariance structure \citep{gu2016parallel}. However, these methods do not incorporate the information of mesh node locations into their models, making it challenging to provide predictions at \textit{any} coordinates in the domain of interest beyond the mesh coordinates. 
Another idea is to perform dimension reduction to approximate the simulation outputs using basis expansion, such as functional principal component analysis \citep{ramsay2005functional} or Karhunen-Lo\`eve expansion \citep{karhunen1947ueber,loeve1955probability}, and then fit GP models on the coefficients, the number of which is usually much less than the number of outputs. The methods adopting this idea include \cite{higdon2008computer}, \cite{rougier2008efficient}, \cite{rougier2009expert}, \cite{marrel2011global}, \cite{mak2017efficient}, and \cite{tan2018gaussian}. Such methods, however, achieve the dimension reduction by a \textit{finite} truncation of the expansion in a function basis, and the approximation error can introduce additional bias to the predictions.

In this paper, we propose a novel emulator, called \textit{mesh-clustered Gaussian process} (\texttt{mcGP}) emulator, for predicting the outputs at many fixed coordinates by incorporating the information of mesh node coordinates. Specifically, instead of fitting separate GPs with different hyperparameters at each node coordinate, the proposed method makes use of the divide-and-conquer idea, which segments the node coordinates into \textit{clusters} with a soft-assignment clustering approach, within each of which, GPs with the \textit{same} hyperparameters are fitted at each coordinate. In particular, the Dirichlet process (DP) prior is adopted here, facilitating a flexible clustering structure for the proposed mixture model without the need for specifying a fixed number of clusters. In addition, a basis expansion representation is employed for mesh-based numerical solutions and the GPs are used to model the coefficients. This approach enables us to make predictions across the entire spatial domain, extending beyond the mesh coordinates. Note that, this basis expansion \textit{does not} perform dimension reduction, so no additional bias will be introduced to our predictions.
Importantly, given such a sophisticated mixture model, the proposed method can be fitted efficiently by adopting the variational Bayesian inference method \citep{jordan1999introduction,wainwright2008graphical}, which provides an analytical approximation to the posterior distribution of the latent variables and parameters, facilitating faster computation and scalability when compared with traditional approaches like Markov chain Monte Carlo (MCMC).



In addition to efficient emulation, our method also provides two important features. First, our method enables efficient uncertainty quantification for  emulation with theoretical guarantees.
In particular, we provide the error analysis that not only considers the uncertainty from the emulator given limited training samples, but also accounts for the numerical error arising from the discrete approximation of the continuous domain through a mesh configuration in numerical methods for solving PDEs. Second, by revealing the clustering structures, the proposed method provides valuable insights into qualitative features of the resulting dynamics.  Unlike traditional reduced-basis methods for flow simulations in physics and engineering, such as proper orthogonal decomposition (POD) \citep{lumley1967structure},  which are unsupervised learning methods purely based on the flow data, the latent clustering structure by the proposed method is determined by both  inputs and outputs as in \cite{joseph2021supervised} and \cite{sung2019clustered}, which can be used to separate a simulated flow into key instability structures, each with its corresponding spatial features.


It is important to note that recent developments in the field of finite element methods have led to the emergence of the \textit{statistical finite elements} (statFEM) approach, which holds promise in advancing uncertainty quantification and model predictions through a Bayesian statistical construction of finite element methods. See, for example, \cite{duffin2021statistical}, \cite{girolami2021statistical} and \cite{akyildiz2022statistical}. Specifically, 
statFEM introduces a hierarchical statistical model to handle uncertainties in data, mathematical models, and finite element discretization. It decomposes data into three components---finite element, model misspecification, and noise---each treated as a random variable with a corresponding prior probability density. 
In contrast, our focus in this paper is primarily on the development of an efficient surrogate model, providing a faster alternative to costly finite element simulations.

The rest of this article is organized as follows. Section \ref{sec:review} provides a brief overview of PDE boundary value problems and mesh-based numerical methods. The proposed model is introduced in Section \ref{secMethod}, and its error analysis is derived in Section \ref{secError}. Real examples are demonstrated in Section \ref{secnum}. Section \ref{secconclu} concludes with directions for future work. Theoretical proofs, and the \textsf{R} \citep{R2018} code for reproducing the numerical results, are provided in Supplementary Materials.

\section{PDE boundary value problems and mesh-based numerical methods}\label{sec:review}

\begin{figure}[h]
    \centering
    \includegraphics[width=0.7\textwidth]{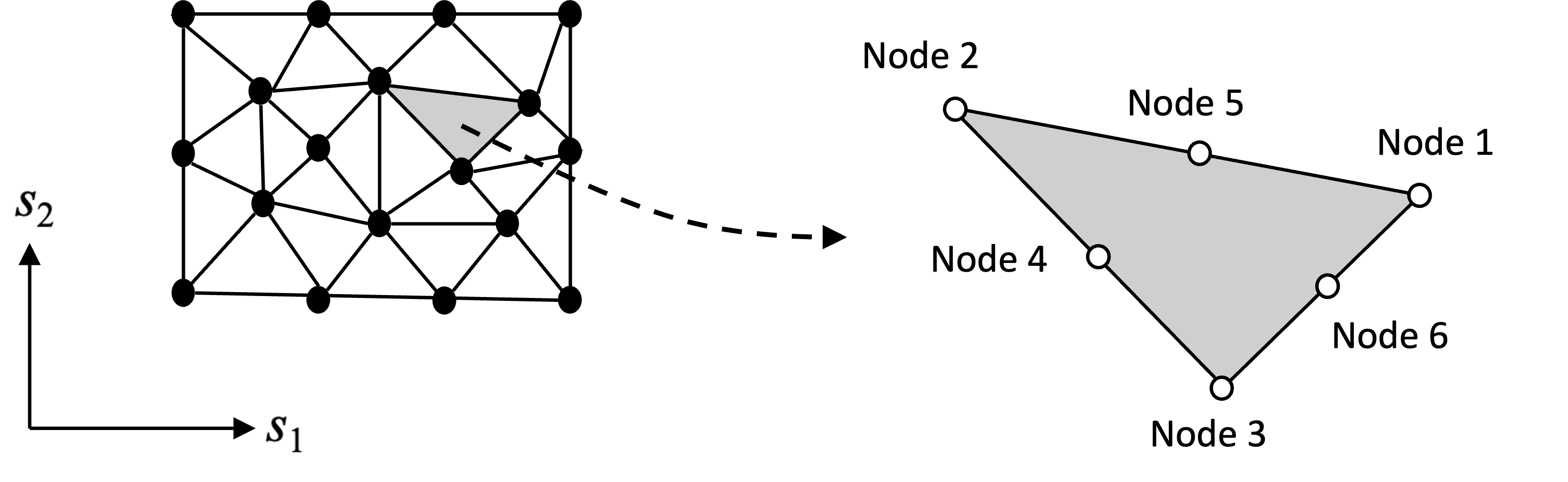}
    \caption{Introduction of finite element method; similar to one from \cite{Tekkaya2019}.}
    \label{fig:femintroduction}
\end{figure}

PDEs are an essential tool in the description of complex systems drawing from scientific principles. A PDE boundary value problem typically can be expressed by a set of partial differential equations with boundary conditions, and the solutions  are often dependent on certain inputs, denoted as $\mathbf{x}\in \chi\subset \RR^p$,   typically representing  parameters in the equations and boundary conditions, where $\chi$ is assumed to be compact and convex. Specifically, a PDE boundary value problem can be written in the form of
\begin{equation}\label{PDEsys}
    \begin{cases}\mathcal{L}(u(\mathbf{s});\mathbf{x}) = f(\mathbf{s};\mathbf{x}),  & \mathbf{s}\in \Omega\\
    \mathcal{G}(u(\mathbf{s});\mathbf{x}) = g(\mathbf{s};\mathbf{x}), & \mathbf{s}\in \partial \Omega,\end{cases}
\end{equation}
where $\Omega$ is a compact domain in $\RR^d$ with a Lipschitz boundary denoted by $\partial \Omega$, $\mathcal{L}(\cdot;\mathbf{x})$ and $\mathcal{G}(\cdot;\mathbf{x})$ are differential operators on $\Omega$ and $\partial \Omega$ given the input $\mathbf{x}$, respectively, $f(\cdot;\mathbf{x})$ and $g(\cdot;\mathbf{x})$ are two known functions on $\Omega$ and $\partial \Omega$ with the input $\mathbf{x}$, and $u(\mathbf{s})$ is the solution to the partial differential equations \eqref{PDEsys}. Given the input $\mathbf{x}$, the exact solution to the equations when uniqueness holds, denoted by $u_0(\mathbf{s};\mathbf{x})$, usually cannot be written explicitly. Instead, numerical methods are used to approximate the exact solution, for which  mesh-based numerical methods are most widely used, including the finite difference method (FDM), finite volume method (FVM), and finite element method (FEM) \citep{brenner2007fem,Tekkaya2019}. The extensions to \textit{mesh-free} methods, such as the collocation method \citep{golberg1999some,wendland1998numerical,fasshauer1999solving,wendland1999meshless,fasshauer1996solving}, are straightforward and will be discussed in the remarks. Specifically, mesh-based numerical methods subdivide a large system into smaller, simpler parts by a particular space discretization in the space dimensions, which is implemented by the construction of a \textit{mesh} of the object. Figure \ref{fig:femintroduction} demonstrates a mesh specification for a 2-dimensional FEM problem, in which the triangular elements connect all characteristic points (called \textit{nodes}) that lie on their circumference, and these connections are mathematically expressed through a set of functions called \textit{shape functions}.

Suppose that there are $N$ nodes in this mesh-based numercial method, the coordinates of which are denoted by $\mathbf{S}_N=(\mathbf{s}_1,\mathbf{s}_2,\ldots,\mathbf{s}_N)$, where $\mathbf{s}_j\in\Omega\cup\partial\Omega$. We assume that $\mathbf{S}_N$ is fixed  across different inputs $\mathbf{x}$. In the case of an adaptive mesh, where $\mathbf{S}_N$ may vary across different inputs, we explore potential methods in Section \ref{secconclu}. The numerical solutions can be expressed as 
\begin{align}\label{PDEnum}
    u_{N}(\mathbf{s};\mathbf{x}) = \sum_{j=1}^N \beta_j(\mathbf{x})v_j(\mathbf{s}),\end{align}
where $v_j(\mathbf{s})$ is the (given) shape function depending on the discretization, and $\beta_j(\mathbf{x})$ is the corresponding coefficient, which is independent of $\mathbf{s}$. The shape function has the Kronecker Delta property, that is, $v_j(\mathbf{s}_i)=1$ if $i=j$ and $v_j(\mathbf{s}_i)=0$ if $i\neq j$. This ensures that the function interpolates the solution at the mesh nodes, i.e., $u_{N}(\mathbf{s}_j;\mathbf{x})=\beta_j(\mathbf{x})$ for $j=1,\ldots,N$. In other words, the solution of the PDE at the mesh node $\mathbf{s}_j$ is equal to $\beta_j(\mathbf{x})$. As an example, for a triangular element with six nodes as in Figure \ref{fig:femintroduction}, the \textit{quadratic} shape functions, which will be adopted in our later implementation in Section \ref{secnum}, can be defined as:
\begin{equation}\label{eq:quadraticshape}
\begin{gathered}
v_1(\mathbf{s})=\xi_1(\mathbf{s})(2\xi_1(\mathbf{s})-1),\quad v_2(\mathbf{s})=\xi_2(\mathbf{s})(2\xi_2(\mathbf{s})-1),\quad v_3(\mathbf{s})=\xi_3(\mathbf{s})(2\xi_3(\mathbf{s})-1),\\
v_4(\mathbf{s})=4\xi_1(\mathbf{s})\xi_2(\mathbf{s}),\quad v_5(\mathbf{s})=4\xi_2(\mathbf{s})\xi_3(\mathbf{s}),\quad v_6(\mathbf{s})=4\xi_1(\mathbf{s})\xi_3(\mathbf{s}),
\end{gathered}
\end{equation}
where $\xi_1,\xi_2,$ and $\xi_3$ represent the barycentric coordinates associated with the three vertices (Nodes 1, 2, and 3, respectively), which are given by
$
\xi_j(\mathbf{s})=a_j+b_js_1+c_js_2,
$ where $a_j,b_j,$ and $c_j$ are determined based on the geometry and orientation of the triangle, and ensure that the shape functions satisfy the property that they are equal to 1 at the corresponding nodes and equal to 0 at the other nodes. This formulation can be easily extended to multiple triangular elements; we refer more detailed information and variations of shape functions to \cite{bathe2006finite}.

The coefficients are obtained by solving a linear system,
\begin{align}\label{PDElinearapp}
    L(\mathbf{x}) \bm{\beta}_{N}(\mathbf{x})=b(\mathbf{x}),
\end{align}
where $\bm{\beta}_{N}(\mathbf{x}) = (\beta_{1}(\mathbf{x}),\ldots,\beta_{N}(\mathbf{x}))^T$ is a vector of the coefficients, and $L(\mathbf{x})\in \RR^{N\times N}$ is the \textit{stiffness matrix} and $b(\mathbf{x})\in \RR^{N}$ is the \textit{load vector}, both of which are determined by the numerical method.


The main challenge of mesh-based numerical methods is that directly solving the PDE boundary value problem for any input $\mathbf{x}$, or equivalently, solving the linear system \eqref{PDElinearapp}, can be computationally demanding (e.g., the high-fidelity simulation in \cite{mak2017efficient}). Thus, an efficient emulator that can approximate the solution, $u_{N}(\mathbf{s};\mathbf{x})$, for any $\mathbf{s}\in\Omega,\mathbf{x}\in\chi$, is called for.

\section{Mesh-clustered Gaussian process (mcGP) emulator}\label{secMethod}

From \eqref{PDEnum}, since emulating $u_{N}(\mathbf{s};\mathbf{x})$ is equivalent to emulating $\{\beta_j(\mathbf{x})\}^N_{j=1}$ (because $v_j$ is a known function), we aim to build an efficient emulator that approximates $\bm{\beta}_{N}(\mathbf{x}):=\{\beta_j(\mathbf{x})\}^N_{j=1}$ for any $\mathbf{x}\in\chi$. Suppose that $n$ computer simulations with the inputs, $\mathbf{x}_1,\ldots,\mathbf{x}_n$, are conducted,  and their corresponding  solutions at the $N$ nodes are $\{\bm{\beta}_N(\mathbf{x}_i)\}^n_{i=1}$. Clearly, this is a multi-output regression problem, because for each input $\mathbf{x}_i$, the output $\bm{\beta}_N(\mathbf{x}_i)$ returns a vector of size $N$, where $N$ can be fairly large. To this end, we propose an efficient emulator that couples over clusters of Gaussian process (GP) emulators, with the aid of the mesh specification to find the clustering structure.

To begin, we briefly introduce the GP emulator in the following subsection.

\subsection{Gaussian process (GP) emulator}\label{subsecGPemulator}
A GP is a popular tool for building an emulator for computer experiments \citep{santner2013design,gramacy2020surrogates} due to its flexibility and the
capability of uncertainty quantification through the predictive distribution. Specifically, suppose that we aim to emulate the single output $\beta_j(\mathbf{x})$, then the function $\beta_j$ can be assumed to have a GP prior with zero mean and a positive-definite covariance function, $K_j(\cdot,\cdot):\RR^p\times\RR^p\rightarrow\RR$. The covariance function often has the form of $K_j(\cdot,\cdot)=\tau_j^2\Phi_{\bm{\theta}_j}(\cdot,\cdot)$, where $\tau_j^2$ is a positive scale and $\Phi_{\bm{\theta}_j}$ is a positive-definite correlation function that depends on some hyperparameters $\bm{\theta}_j$. We denote such a GP as
$$
\beta_j(\mathbf{x})|\tau^2_j,\bm{\theta}_j\sim\mathcal{GP}(0,\tau^2_j\Phi_{\bm{\theta}_j}(\mathbf{x},\mathbf{x}')).
$$
The GP assumes that the random vector $\mathbf{b}_j=(\beta_j(\mathbf{x}_1),\ldots,\beta_j(\mathbf{x}_n))^T$ follows an $n$-dimensional multivariate normal distribution, $\mathcal{N}_n(0,\tau_j^2\Phi_{\bm{\theta}_j}(\mathbf{X}_n,\mathbf{X}_n)),$
where $\mathbf{X}_n=(\mathbf{x}_1,\ldots,\mathbf{x}_n)$ and $\Phi_{\bm{\theta}_j}(\mathbf{X}_n,\mathbf{X}_n)$ is an $n\times n$ matrix with $(\Phi_{\bm{\theta}_j}(\mathbf{X}_n,\mathbf{X}_n))_{i,k}=\Phi_{\bm{\theta}_j}(\mathbf{x}_i,\mathbf{x}_k)$. To emulate the $N$ outputs $\bm{\beta}_N(\mathbf{x})=(\beta_1(\mathbf{x}),\ldots,\beta_N(\mathbf{x}))^T$, we assume that the GPs are independent, implying that the $N$ random vectors, $\mathbf{b}_1,\ldots,\mathbf{b}_N$, are independent.


For a new input $\mathbf{x}$, it can be shown that the posterior predictive distribution is a normal distribution with the mean \begin{eqnarray}\label{mean}
&\mathbb{E}[\beta_j(\mathbf{x})|\mathbf{b}_j,\tau^2_j,\bm{\theta}_j]=\Phi_{\bm{\theta}_j}(\mathbf{x},\mathbf{X}_n) \Phi_{\bm{\theta}_j}(\mathbf{X}_n,\mathbf{X}_n)^{-1}\mathbf{b}_j
\end{eqnarray}
and the variance
\begin{eqnarray}\label{var}
&\mathbb{V}[\beta_j(\mathbf{x})|\mathbf{b}_j,\tau^2_j,\bm{\theta}_j]=\tau_j^2(1-\Phi_{\bm{\theta}_j}(\mathbf{x},\mathbf{X}_n) \Phi_{\bm{\theta}_j}(\mathbf{X}_n,\mathbf{X}_n)^{-1}\Phi_{\bm{\theta}_j}(\mathbf{x},\mathbf{X}_n)^T),
\end{eqnarray}
where $\Phi_{\bm{\theta}_j}(\mathbf{x},\mathbf{X}_n)$ is a $1\times n$ matrix with $(\Phi_{\bm{\theta}_j}(\mathbf{x},\mathbf{X}_n))_{1,i}=\Phi_{\bm{\theta}_j}(\mathbf{x},\mathbf{x}_i)$. The posterior predictive mean can be used to predict $\beta_j(\mathbf{x})$ and the posterior predictive variance can be used to quantify the prediction uncertainty.

Two families of correlation functions are widely used in practice, which are the power exponential correlation functions and Mat\'ern correlation function \citep{santner2013design,stein2012interpolation}. For instance, the Mat\'ern correlation function has the form of
\begin{eqnarray}\label{matern}
\Phi_{\bm{\theta}}(\mathbf{x}_i,\mathbf{x}_j)=\frac{1}{\Gamma(\nu)2^{\nu-1}}(2\sqrt{\nu}\|\mathbf{x}_i-\mathbf{x}_j\|_{\bm{\theta}})^\nu B_\nu(2\sqrt{\nu}\|\mathbf{x}_i-\mathbf{x}_j\|_{\bm{\theta}}),
\end{eqnarray}
where $\|a\|_{\bm{\theta}}^2 = \sum_{k=1}^p  (a_k/\theta_k)^2$ with the $p$-dimensional lengthscale hyperparameter $\bm{\theta}=(\theta_1,\ldots,\theta_p)$, $\nu>0$ is the smoothness parameter \citep{cramer2013stationary}, and $B_\nu$ is the modified Bessel function of the second kind. 

\subsection{Model specification}\label{subsecmgGPemulator}
While it is possible to model multivariate GPs, instead of independent GPs, using a separable covariance function \citep{NIPS2007_66368270,qian2008gaussian} or a nonseparable covariate function \citep{fricker2013multivariate,svenson2016multiobjective}, fitting these models can be computationally prohibitive when $N$ is large. In addition, recent studies \citep{zhang2015doesn,kleijnen2014multivariate,li2020prediction} have shown that such multivariate GPs could actually yield worse prediction accuracy than the independent (univariate) GPs described in Section \ref{subsecGPemulator}. On the other hand, \cite{gu2016parallel} considers independent GPs that share the same lengthscale hyperparameters over the $N$ outputs, i.e., $\bm{\theta}_1=\cdots=\bm{\theta}_N$, but assume different $\tau_j$'s over the $N$ outputs. 

From the perspective of statistical learning, independent univariate GPs sharing the same hyperparameters over the $N$ outputs could be underparametrized, while the one sharing different hyperparameters could be overparametrized. To this end, we propose a flexible model that serves as a compromise between the two models:
\begin{align}\label{eq:oriDP}
    \beta_j(\mathbf{x})|\tau^2_j,\bm{\theta}_j&\sim\mathcal{GP}(0,\tau^2_j\Phi_{\bm{\theta}_j}(\mathbf{x},\mathbf{x}'))\quad\text{for}\quad j=1,\ldots,N,\nonumber\\
    \tau^2_j, \bm{\theta}_j&\sim G,\\
    G&\sim\mathcal{DP}(H,\alpha_0),\nonumber
\end{align}
where $\mathcal{DP}(H,\alpha_0)$ denotes a Dirichlet process (DP) prior \citep{ferguson1973bayesian} with a positive real scalar $\alpha_0$ and $H$ being a distribution over $\tau^2_j$ and $\bm{\theta}_j$. The parameter $\alpha_0$ is often called  \textit{concentration} parameter. The smaller $\alpha_0$ yields more concentrated distributions. The DP prior is a Bayesian nonparametric model and is a popular tool for developing mixture models, which are often called \textit{infinite} mixture models, because such mixture models have a countably infinite number of mixture components; therefore, these models do not require to pre-specify a fixed number of mixture components, which can be difficult to determine in practice. 
A DP can be constructively defined by the \textit{stick-breaking} construction \citep{sethuraman1994constructive}, by which \eqref{eq:oriDP} can be equivalently expressed as 
\begin{align}
   \beta_j(\mathbf{x})|z_j=k,\tau^2_k,\bm{\theta}_k&\sim\mathcal{GP}(0,\tau^2_k\Phi_{\bm{\theta}_k}(\mathbf{x},\mathbf{x}'))&\text{for}&\quad j=1,\ldots,N,\label{eq:mixtureGP}\\
    z_j|\gamma_1,\ldots,\gamma_{\infty}&\stackrel{\text{iid}}{\sim}\text{Categorical}(\pi_1,\pi_2,\ldots,\pi_{\infty}) &\text{for}&\quad j=1,\ldots,N,\label{eq:zgivengamma}\\
 \pi_k&=\gamma_k\prod^{k-1}_{l=1}(1-\gamma_l)&\text{for}&\quad k=1,2,\ldots,\infty,\label{eq:pikstick}\\
   \gamma_k&\stackrel{\text{iid}}{\sim}\text{Beta}(1,\alpha_0)\label{eq:gammadist}&\text{for}&\quad k=1,2,\ldots,\infty,\\
    \tau^2_k, \bm{\theta}_k&\stackrel{\text{iid}}{\sim} H,&\text{for}&\quad k=1,2,\ldots,\infty,\nonumber
\end{align}
where ``iid'' denotes independently and identitically distributed,  and ``Categorical'' and ``Beta'' denote a categorical distribution and  a Beta distribution, respectively. From \eqref{eq:mixtureGP}, $z_j$ is a latent variable indicating the assignment of $\beta_j(\mathbf{x})$ to the $k$-th GP having the hyperparameters $\bm{\theta}_k$ and $\tau_k^2$, where the number of GPs is  countably infinite. This forms an infinite mixture of GPs for  multivariate outputs, with the mixing proportion $\text{Pr}(z_j=k)=\pi_k$. The proportion $\pi_k$ is given by \eqref{eq:pikstick}, which provides the \textit{stick-breaking} representation of a DP as $G=\sum^{\infty}_{k=1}\pi_k\delta_{(\tau^2_k,\bm{\theta}_k)}$, where $\delta_{(\tau^2_k,\bm{\theta}_k)}$ is the indicator function whose value is one at location $(\tau^2_k,\bm{\theta}_k)$ and zero elsewhere.  The proportions $\{\pi_k\}^{\infty}_{k=1}$ always sum to one and can be resembling the breaking of a unit-length stick into a countably infinite number of pieces (hence the name). 
That is, a portion of a unit-length stick is broken off according to $\gamma_k$ and assigned to $\pi_k$, so \eqref{eq:pikstick} can  be understood by considering that after the first $k-1$  
values have been assigned their portions, the length of the remaining stick is $\prod^{k-1}_{l=1}(1-\gamma_l)$, and  this remaining piece is  broken according to $\gamma_k$ and  assigned to $\pi_k$. It is important to note that because the $\pi_k$'s decrease exponentially quickly, only a small number of components will be used to model the data a priori, which allows the number of clusters to  be automatically determined. More details about DP applications can be found in \cite{neal1992bayesian}, \cite{lo1984class}, and \cite{rasmussen1999infinite}.

We further let the latent indicator variable be \textit{mesh-dependent}, implying that the clustering structure is determined by the mesh coordinates. Specifically, assume that a node coordinate $\mathbf{s}$ for the cluster $k$ follows a $d$-dimensional multivariate normal distribution,
\begin{equation}\label{eq:nodedist}
\mathbf{s}|z=k\sim\mathcal{N}_d(\bm{\mu}_k,\bm{\Sigma}_k^{-1}),
\end{equation}
where $\bm{\mu}_k$ is the mean and $\bm{\Sigma}_k$ is the precision matrix, and their priors are a multivariate normal distribution and a Wishart distribution, respectively, that is,
\begin{equation}\label{eq:musigmadist}
\bm{\mu}_k\sim\mathcal{N}_d(\bm{\mu}_0,\bm{\Sigma}_0^{-1}),\quad\bm{\Sigma}_k\sim\mathcal{W}(\mathbf{W}_0,\kappa_0),
\end{equation}
where $\bm{\mu}_0,\bm{\Sigma}_0,\mathbf{W}_0$ and $\kappa_0$ are fixed hyperparameters. 
Unlike a conditional model where  $z|\mathbf{s}$ is employed to model the dependence of $z$ and $\mathbf{s}$, the model \eqref{eq:nodedist} is a \textit{generative} model \citep{ng2001discriminative}, which assumes that the mesh coordinates $\{\mathbf{s}_j\}^N_{j=1}$, while observed and fixed, are treated as instances or realizations generated from the underlying distribution $\mathbf{s} | z$. 
This conceptualization allows us to incorporate uncertainty related to distinct instances of mesh coordinates within the same cluster, and offers a consistent way of specifying each component's responsibility for a given node coordinate \citep{meeds2005alternative,sun2010variational}. It is also worth noting that the multivariate normal assumption for $\mathbf{s}|z$ aligns with a well-known classification method---quadratic discriminant analysis (QDA), implying that mesh nodes will be categorized into distinct classes/clusters based on this model choice.

\begin{remark}
Although the latent indicator variable is mesh-dependent particularly for mesh-based numerical methods, the idea can also be extended to mesh-free numerical methods, such as the collocation method \citep{wendland1998numerical}, where the basis function $v_j(\mathbf{s})$ becomes radial basis functions with some given knot locations and the knots can then be clustered in a similar manner. In addition, the idea can be naturally extended to other applications than solving PDEs that may have different dependence structures, such as the spatially-related dependence structure between proximate sites in \cite{dahl2019grouped} and the network-related dependence structure in \cite{wilson2011gaussian}.
\end{remark}

\begin{figure}[t]
    \centering
    \includegraphics[width=0.7\linewidth]{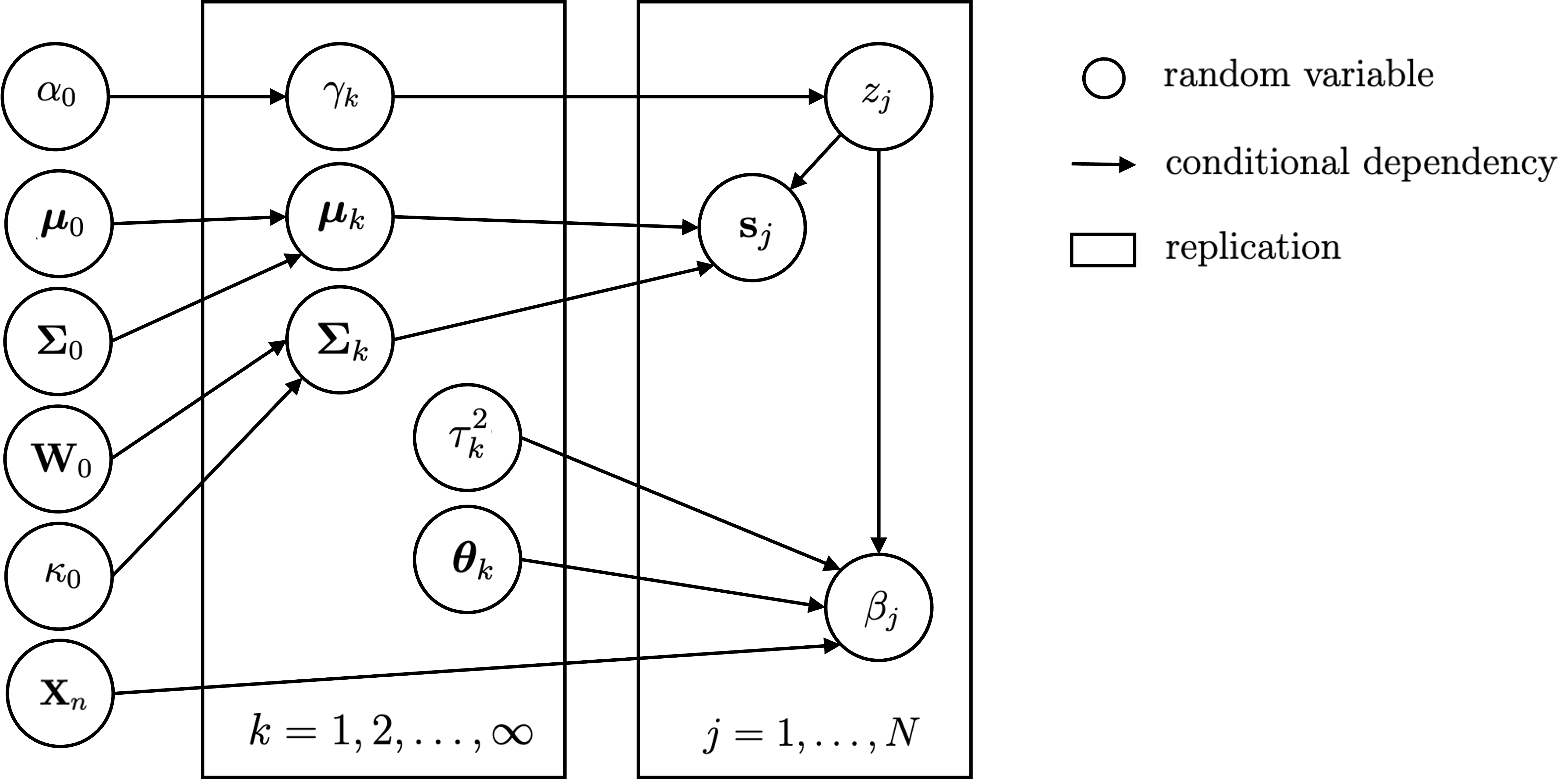}
    \caption{Graphical model representation of \texttt{mcGP}.}
    \label{fig:graphicalmgGP}
\end{figure}

Combining the above models, a graphical representation of the proposed model is given in Figure \ref{fig:graphicalmgGP}. It should be noted that the proposed \texttt{mcGP} is intrinsically different from the infinite mixture of
GPs in \cite{rasmussen2001infinite}, \cite{meeds2005alternative} and \cite{sun2010variational}. In particular, the mixture model therein focuses on dividing the input space of $\mathbf{x}$  to address both the problems of computational complexity and stationary assumption for a univariate GP, whereas our proposed method addresses multi-output regression problems by  dividing the mesh node coordinates, $\{\mathbf{s}_j\}^N_{j=1}$, into regions, within which separate univariate GPs with the same hyperparameters make predictions. 

\subsection{Parameter estimation: Variational inference}\label{secParameter}
Although Markov chain Monte Carlo (MCMC) methods, such as Gibbs sampling, can be naturally used to draw the posterior distribution of hidden variables for DP mixture models (see, e.g., \cite{rasmussen1999infinite,neal2000markov,rasmussen2001infinite}), they are often computationally demanding.  Therefore, variational inference (VI) \citep{jordan1999introduction,ganguly2021introduction}  is adopted here to approximate the posterior, which leads to faster computation and efficient scalability.

Denote the parameters $\tilde{\bm{\gamma}}=(\gamma_1,\ldots,\gamma_{\infty}),\tilde{\bm{\mu}}=(\bm{\mu}_1,\ldots,\bm{\mu}_{\infty}),\tilde{\bm{\Sigma}}=(\bm{\Sigma}_1,\ldots,\bm{\Sigma}_{\infty})$ and the latent variable $\tilde{\mathbf{z}}=(z_1,\ldots,z_N)$, and denote the hyperparameters $\tilde{\bm{\tau}}^2=(\tau^2_1,\ldots,\tau^2_{\infty})$ and $\tilde{\bm{\theta}}=(\bm{\theta}_1,\ldots,\bm{\theta}_{\infty})$. Similar to \cite{sun2010variational}, the hyperparameters $\tilde{\bm{\tau}}^2$ and $\tilde{\bm{\theta}}$ will be estimated via variational expectation maximization (EM). We first develop the posterior of the  hidden variable vector, which is denoted by $\bm{\phi}:=(\tilde{\bm{\gamma}},\tilde{\bm{\mu}},\tilde{\bm{\Sigma}},\tilde{\mathbf{z}})$. 

Denote $\mathcal{D}$ as the observational data,  $\mathcal{D}=\{\beta_j(\mathbf{x}_1),\ldots,\beta_j(\mathbf{x}_n),\mathbf{s}_j\}^N_{j=1}=\{\mathbf{b}_j,\mathbf{s}_j\}^N_{j=1}$. By the graphical model representation in Figure \ref{fig:graphicalmgGP}, the joint distribution is
\begin{align}\label{eq:evidence}
p(\mathcal{D},\bm{\phi})&=p(\mathcal{D}|\bm{\phi})p(\bm{\phi})=\left(\prod^N_{j=1}p(\mathbf{b}_j|z_j)p(\mathbf{s}_j,z_j|\tilde{\bm{\mu}},\tilde{\bm{\Sigma}},\tilde{\bm{\gamma}})\right)\times \left(\prod^{\infty}_{k=1}p(\gamma_k)p(\bm{\mu}_k)p(\bm{\Sigma}_k)\right)\nonumber\\
    &=\left(\prod^N_{j=1}p(\mathbf{b}_j|z_j)p(\mathbf{s}_j|z_j,\tilde{\bm{\mu}},\tilde{\bm{\Sigma}})p(z_j|\tilde{\bm{\gamma}})\right)\times \left(\prod^{\infty}_{k=1}p(\gamma_k)p(\bm{\mu})p(\bm{\Sigma}_k)\right),
\end{align}
where $p(\mathbf{b}_j|z_j)$ is the probability density function (pdf) of the $n$-dimensional multivariate normal distribution from \eqref{eq:mixtureGP}, i.e., $\mathbf{b}_j|z_j=k\sim \mathcal{N}_n(0,\tau_k^2\Phi_{\bm{\theta}_k}(\mathbf{X}_n,\mathbf{X}_n))$,  $p(\mathbf{s}_j|z_j,\tilde{\bm{\mu}},\tilde{\bm{\Sigma}})$ is the pdf of the multivariate normal distribution from \eqref{eq:nodedist}, $p(\gamma_k)$ is the beta distribution from \eqref{eq:gammadist}, and $p(\bm{\mu}_k)$ and $p(\bm{\Sigma}_k)$ are the multivariate normal distribution and the Wishart distribution from \eqref{eq:musigmadist}, respectively.

Clearly, the posterior, $p(\bm{\phi}|\mathcal{D})=p(\mathcal{D},\bm{\phi})/p(\mathcal{D})$, has a complex probability density which cannot be represented in a closed form. To this end, we apply VI to provide an analytical approximation to $p(\bm{\phi}|\mathcal{D})$. Specifically, VI finds a distribution that is restricted to belong to a family of distributions of simpler forms, denoted by $q(\bm{\phi})$, such that $q(\bm{\phi})\approx p(\bm{\phi}|\mathcal{D})$, which is called  \textit{variational distribution}. This can be done by finding the variational distribution that minimizes the Kullback-Leibler (KL) divergence of $q(\bm{\phi})$ from $p(\bm{\phi}|\mathcal{D})$:
\begin{align*}
\text{KL}(q(\bm{\phi})||p(\bm{\phi}|\mathcal{D}))&=\mathbb{E}_q[\log(q(\bm{\phi}))]-\mathbb{E}_q[\log p(\bm{\phi}|\mathcal{D})]\\
&=\mathbb{E}_q[\log(q(\bm{\phi}))]-\mathbb{E}_q[\log p(\mathcal{D},\bm{\phi})]+\log p(\mathcal{D}),
\end{align*}
which is equivalent to maximizing
\begin{equation}\label{eq:elbo}
    \text{ELBO}(q)=\mathbb{E}_q[\log p(\mathcal{D},\bm{\phi})]-\mathbb{E}_q[\log(q(\bm{\phi}))]
\end{equation}
because $\log p(\mathcal{D})=\text{KL}(q(\bm{\phi})||p(\bm{\phi}|\mathcal{D}))+\text{ELBO}(q)$ and $\log p(\mathcal{D})$ is fixed with respect to $q(\bm{\phi})$. The function is called \textit{evidence lower bound (ELBO)}, which is a lower bound for the log-evidence of the data (hence the name), that is, $\log p(\mathcal{D},\bm{\phi})\geq \text{ELBO}(q)$ for any $q$. This can be shown by the fact that $\text{KL}(q(\bm{\phi})||p(\bm{\phi}|\mathcal{D}))\geq 0$ for any $q$.


Here we adopt the \textit{mean field approximation} method  (see \cite{blei2017variational} for more details) to formulate the variational distribution $q(\bm{\phi})$, which considers the mean-field variational family where   the variables $\bm{\phi}$ are mutually independent and each governed by a distinct factor in the variational density. In addition, we approximate the posterior DP by a \textit{truncated} stick-breaking representation as in \cite{ishwaran2001gibbs}, \cite{blei2006variational}, and \cite{sun2010variational}. This can be done by setting $q(\gamma_K=1) = 1$ with a fixed value $K$, indicating that the stick is no longer broken after $K-1$ steps. The length of the remainder of the stick is assigned to $\pi_K$, while the lengths of the sticks beyond $K$ become zero. This implies that the mixing proportion $\pi_k=0$ for any $k>K$. Together, the mean field approximation method uses the following factorized variational distribution to approximate $p(\bm{\phi}|\mathcal{D})$: 
\begin{equation*}
    q(\bm{\phi})=\prod^{\infty}_{k=1}q(\gamma_k)q(\bm{\mu}_k)q(\bm{\Sigma}_k)\prod^N_{j=1}q(z_j)=\prod^{K-1}_{k=1}q(\gamma_k)\prod^{K}_{k=1}q(\bm{\mu}_k)q(\bm{\Sigma}_k)\prod^N_{j=1}q(z_j).
\end{equation*}
Given this mean-field variational family, the variational distribution for each $\omega\in\bm{\phi}$ can be optimized by maximizing the ELBO using coordinate ascent variational inference \citep{bishop2006pattern}, which gives the optimal variational distribution: 
\begin{equation}\label{eq:meanapproximation}
\log q(\omega)=\mathbb{E}_{\bm{\phi}\setminus\omega}[\log p(\mathcal{D},\bm{\phi})]+\text{constant},
\end{equation}
where the expectation $\mathbb{E}_{\bm{\phi}\setminus\omega}$ is with respect to the variational distribution over all variables $\bm{\phi}$ except for $\omega$. For example, from \eqref{eq:evidence}, the optimal variational distribution for $\gamma_k$ is
\begin{align*}
\log q(\gamma_k)&=\mathbb{E}_{\bm{\phi}\setminus\gamma_k}[\log p(\mathcal{D},\bm{\phi})]+\text{constant}\\
&=\sum^N_{j=1}\mathbb{E}_{\bm{\phi}\setminus\gamma_k}[\log p(z_j|\tilde{\bm{\gamma}})]+\log p(\gamma_k)+\text{constant}.
\end{align*}
The distribution $p(z_j|\tilde{\bm{\gamma}})$ can be found from \eqref{eq:zgivengamma}. It follows that
\begin{align*}
\mathbb{E}_{\bm{\phi}\setminus\gamma_k}[\log p(z_j|\tilde{\bm{\gamma}})]&=\mathbb{E}_{\bm{\phi}\setminus\gamma_k}[\log \pi_{z_j}]=\mathbb{E}_{\bm{\phi}\setminus\gamma_k}\left[\log \gamma_{z_j}+\sum^{z_j-1}_{l=1}\log (1-\gamma_l)\right]\\
&=\sum^{\infty}_{k'=1}q(z_j=k')\mathbb{E}_{\bm{\phi}\setminus\gamma_k}[\log \gamma_{k'}]+\sum^{\infty}_{k'=1}q(z_j=k')\sum^{k'-1}_{l=1}\mathbb{E}_{\bm{\phi}\setminus\gamma_k}[\log (1-\gamma_l)]\\
&=\sum^{\infty}_{k'=1}q(z_j=k')\mathbb{E}_{\bm{\phi}\setminus\gamma_k}[\log \gamma_{k'}]+\sum^{\infty}_{l=1}\sum^{\infty}_{k'=l+1}q(z_j=k')\mathbb{E}_{\bm{\phi}\setminus\gamma_k}[\log (1-\gamma_l)]\\
&=q(z_j=k)\log \gamma_{k}+q(z_j>k)\log (1-\gamma_{k})+\text{constant},
\end{align*}
where the third equation is based on the variational distribution of $z_j$, i.e., $q(z_j)$. The fourth equation changes the order of summation, and the last equation holds because the terms $\mathbb{E}_{\bm{\phi}\setminus\gamma_k}[\log \gamma_{k'}]$ and $\mathbb{E}_{\bm{\phi}\setminus\gamma_k}[\log (1-\gamma_{k'})]$ are constant with respect to $\gamma_k$ when $k'\neq k$. Since $p(\gamma_k)$ follows a Beta distribution as in \eqref{eq:gammadist}, we have
\begin{align*}
    \log q(\gamma_k)&=\sum^{N}_{j=1}q(z_j=k)\log \gamma_{k}+q(z_j>k)\log (1-\gamma_{k})+(\alpha_0-1)\log(1-\gamma_k)+\text{constant}\\
    &=\left(\sum^{N}_{j=1}q(z_j=k)\right)\log \gamma_{k}+\left(\sum^{N}_{j=1}q(z_j>k)+\alpha_0-1\right)\log (1-\gamma_{k})+\text{constant},
\end{align*}
which implies that $q(\gamma_k)$ follows $\text{Beta}(\sum^{N}_{j=1}q(z_j=k)+1,\sum^{N}_{j=1}q(z_j>k)+\alpha_0)$.

The optimal variational distributions for the remaining variables $\omega\in\bm{\phi}$ can be derived in a similar manner. The specific results are provided in the E-step of Algorithm \ref{alg:variationalEM}, and for detailed derivations, we refer to Supplementary Materials S1. Note that the truncation level $K$ is a variational parameter which can be freely set. Although it is  possible to optimize $K$ with respect to the KL divergence, we hold it fixed as in \cite{blei2006variational} and \cite{sun2010variational} throughout this paper.

Given the  optimal variational distributions, the hyperparameters of the GPs, $\bm{\theta}_k$ and $\tau^2_k$, can be estimated by maximizing the expected log-likelihood with respect to the approximated distributions, which is called \textit{variational EM} in the literature \citep{blei2017variational}. Specifically, since $\mathbf{b}_j|z_j=k\sim \mathcal{N}_n(0,\tau_k^2\Phi_{\bm{\theta}_k}(\mathbf{X}_n,\mathbf{X}_n))$, the estimates, denoted by $\hat{\bm{\theta}}_k$ and $\hat{\tau}^2_k$, can be solved by maximizing
\begin{align}\label{eq:objMstep}
\mathbb{E}_q&[\log p(\mathcal{D},\bm{\phi})]=\text{constant}\nonumber\\
&-\frac{1}{2}\left(\sum^N_{j=1}\sum^K_{k=1}q(z_j=k)\left[n\log\tau^2_k+\log|\Phi_{\bm{\theta}_k}(\mathbf{X}_n,\mathbf{X}_n)|+\frac{1}{\tau_k^2}\mathbf{b}^T_j\Phi_{\bm{\theta}_k}(\mathbf{X}_n,\mathbf{X}_n)^{-1}\mathbf{b}_j\right]\right)
\end{align}
with respect to $\bm{\theta}_k$ and $\tau^2_k$. The estimators are given in the M-step of Algorithm \ref{alg:variationalEM}, and the detailed derivations are provided in Supplementary Materials S1. The E- and M-steps repeat iteratively by updating the parameters of the variational distributions until the ELBO converges, where the explicit form of ELBO (derived from \eqref{eq:elbo})  is provided in Supplementary Materials S1. In total, each iteration going through all the observations would take at most $O(KNn^3)$, which is linear with respect to the number of mesh nodes, $N$. We have developed an open-source \textsf{R} package \texttt{mcGP}, enabling the implementation of the variational EM algorithm described in Algorithm \ref{alg:variationalEM}.


\begin{algorithm}[!b]
\setstretch{0.95}
\caption{Variational expectation maximization for parameter estimation}
\label{alg:variationalEM}
\begin{algorithmic}[1]
\stb Set the truncation level $K>1$ and the hyperparameters $\alpha_0,\bm{\mu}_0,\bm{\Sigma}_0,\mathbf{W}_0$, and $\kappa_0$.
\Repeat\\
\quad \underline{E-step}: variational distributions  $q(\gamma_k),q(\bm{\mu}_k),q(\bm{\Sigma}_k)$ and $q(z_j)$.
\stb $\gamma_k\sim\text{Beta}(a_k,b_k)$ for $k=1,\ldots,K-1$, where
$$
a_k=\sum^N_{j=1}q(z_j=k)+1,\quad b_k=\sum^N_{j=1}q(z_j>k)+\alpha_0.
$$
\stb $\bm{\mu}_k\sim\mathcal{N}_d((\bm{\Sigma}_0+\mathbf{R}_{k2})^{-1}(\bm{\Sigma}_0\bm{\mu}_0+\mathbf{R}_{k1}),(\bm{\Sigma}_0+\mathbf{R}_{k2})^{-1})$ for $k=1,\ldots,K$, where
$$
\mathbf{R}_{k1}=\sum^N_{j=1}q(z_j=k)\mathbb{E}_q[\bm{\Sigma}_k]\mathbf{s}_j,\quad\mathbf{R}_{k2}=\sum^N_{j=1}q(z_j=k)\mathbb{E}_q[\bm{\Sigma}_k].
$$
\stb  $\bm{\Sigma}_k\sim\mathcal{W}(\mathbf{W}_k,\kappa_k)$, where $\kappa_k=\kappa_0+\sum^N_{j=1}q(z_j=k)$ and
$$
\mathbf{W}_k^{-1}=\mathbf{W}_0^{-1}+\sum^N_{j=1}q(z_j=k)\mathbb{E}_q[(\mathbf{s}_j-\bm{\mu}_k)(\mathbf{s}_j-\bm{\mu}_k)^T].
$$
\stb $q(z_j=k)=r_{jk}/\sum^K_{k=1}{r_{jk}}$ for $j=1,\ldots,N$ and $k=1,\ldots,K$ with \begin{align*}\log r_{jk}&=\mathbb{E}_q[\log \gamma_k]+\sum^{k-1}_{i=1}\mathbb{E}_q[\log (1-\gamma_i)]+\frac{1}{2}(s_{jk}+t_{jk}), \quad\text{where}\\
s_{jk}&=-d\log(2\pi)+\mathbb{E}_q[\log|\mathbf{\Sigma}_k|]-\mathbb{E}_q[(\mathbf{s}_j-\bm{\mu}_k)^T\mathbf{\Sigma}_k(\mathbf{s}_j-\bm{\mu}_k)]\quad\text{and}\\
t_{jk}&=-n\log(2\pi)-\log\tau_k^2-\log|\Phi_{\bm{\theta}_k}(\mathbf{X}_n,\mathbf{X}_n)|-\frac{1}{\tau_k^2}\mathbf{b}^T_j\Phi_{\bm{\theta}_k}(\mathbf{X}_n,\mathbf{X}_n)^{-1}\mathbf{b}_j.
\end{align*}
\quad \underline{M-step}: estimating the hyperparameters $\bm{\theta}_k$ and $\tau^2_k$.
\stb $\bm{\theta}_k\leftarrow\argmin_{\bm{\theta}_k}\log|\Phi_{\bm{\theta}_k}(\mathbf{X}_n,\mathbf{X}_n)|+n\log\sum^N_{j=1}q(z_j=k)\mathbf{b}^T_j\Phi_{\bm{\theta}_k}(\mathbf{X}_n,\mathbf{X}_n)^{-1}\mathbf{b}_j$
\stb $\tau^2_k\leftarrow\left(\sum^N_{j=1}q(z_j=k)\mathbf{b}^T_j\Phi_{\bm{\theta}_k}(\mathbf{X}_n,\mathbf{X}_n)^{-1}\mathbf{b}_j\right)/\left(n\sum^N_{j=1}q(z_j=k)\right)$
\Until{ELBO converges}
\stb \Return $q(z_j=k),\bm{\theta}_k$ and $\tau^2_k$ for each $k$ and $j$.
\normalsize
\end{algorithmic}
\end{algorithm}

\subsection{Prediction}\label{sec:prediction}

For unknown $\mathbf{x}\in\chi$, the predictive posterior distribution of  $\beta_j(\mathbf{x})$ can be constructed as:
\begin{align*}
    p(\beta_j(\mathbf{x})|\mathcal{D},\bm{\phi},\{\hat{\bm{\theta}}_k,\hat{\tau}^2_k\}^K_{k=1})=&\sum^K_{k=1}p(z_j=k|\mathcal{D},\bm{\phi})p(\beta_j(\mathbf{x})|z_j=k,\mathcal{D},\hat{\bm{\theta}}_k,\hat{\tau}^2_k)\nonumber\\
    \approx&\sum^K_{k=1}q(z_j=k)p(\beta_j(\mathbf{x})|z_j=k,\mathcal{D},\hat{\bm{\theta}}_k,\hat{\tau}^2_k),
\end{align*}
where the approximation replaces $p(z_j=k|\mathcal{D},\bm{\phi})$ with its variational distribution $q(z_j=k)$ as in \cite{sun2010variational}, and $p(\beta_j(\mathbf{x})|z_j=k,\mathcal{D},\{\hat{\bm{\theta}}_k,\hat{\tau}^2_k\}^K_{k=1})$ is the pdf of a normal distribution with the mean \eqref{mean} and the variance \eqref{var} by replacing $\bm{\theta}_j$ and $\tau_j^2$ with  $\hat{\bm{\theta}}_k$ and $\hat{\tau}^2_k$, respectively. Thus, since $u_{N}(\mathbf{s};\mathbf{x}) = \sum_{j=1}^N \beta_j(\mathbf{x})v_j(\mathbf{s})$, the posterior predictive mean of $u_N(\mathbf{s};\mathbf{x})$ for any $\mathbf{s}\in\Omega$ and  $\mathbf{x}\in\chi$, can be approximated by
\begin{equation}\label{eq:mgGPemulator}
\hat{u}_N(\mathbf{s};\mathbf{x}):=\sum^N_{j=1}\sum^K_{k=1}q(z_j=k)v_j(\mathbf{s})\Gamma_{\hat{\bm{\theta}}_k}(\mathbf{x},\mathbf{X}_n) \mathbf{b}_j,
\end{equation}
where $\Gamma_{\hat{\bm{\theta}}_k}(\mathbf{x},\mathbf{X}_n)=\Phi_{\hat{\bm{\theta}}_k}(\mathbf{x},\mathbf{X}_n) \Phi_{\hat{\bm{\theta}}_k}(\mathbf{X}_n,\mathbf{X}_n)^{-1}$, and the posterior predictive variance is
\begin{equation}\label{eq:postvar}
\begin{split}
\sum^N_{j=1}v_j(\mathbf{s})^2\Biggl(\sum^K_{k=1}q(z_j=k)\Biggl[\hat{\tau}_k^2(1-\Gamma_{\hat{\bm{\theta}}_k}(\mathbf{x},\mathbf{X}_n)\Phi_{\hat{\bm{\theta}}_k}(\mathbf{x},\mathbf{X}_n)^T)+
(\Gamma_{\hat{\bm{\theta}}_k}(\mathbf{x},\mathbf{X}_n)\mathbf{b}_j)^2\Biggr]\\-\left(\sum^K_{k=1}q(z_j=k)\Gamma_{\hat{\bm{\theta}}_k}(\mathbf{x},\mathbf{X}_n) \mathbf{b}_j\right)^2\Biggr).
\end{split}
\end{equation}
The derivation of \eqref{eq:postvar} is provided in Supplementary Materials S2. The posterior predictive mean $\hat{u}_N(\mathbf{s};\mathbf{x})$ of \eqref{eq:mgGPemulator} is used to predict $u_N(\mathbf{s};\mathbf{x})$, and its error analysis is studied in the next section.

\section{Error analysis of mcGP emulator}\label{secError}
The error analysis is crucial for understanding the uncertainty of the emulator.  By the triangle inequality, the norm of the difference between the posterior predictive mean $\hat{u}_N(\mathbf{s};\mathbf{x})$ and  true solution can be decomposed into \textit{evaluation}  and  \textit{emulation} components:
$$ \|u_0-\hat{u}_N\|_{L_2(\Omega,\chi)}\leq\underbrace{\|u_0-u_N\|_{L_2(\Omega,\chi)}}_\text{numerical error}+\underbrace{\|u_N-\hat{u}_{N}\|_{L_2(\Omega,\chi)}}_\text{emulation error},$$
where the $L_2$-norm is defined as $\|f\|_{L_2(\Omega,\chi)}=(\int_{\Omega}\int_{\chi}f(\mathbf{s};\mathbf{x})^2\text{d}\mathbf{s}\text{d}\mathbf{x})^{1/2}$. 
The numerical error measures the discrepancy between
the numerical solution $u_N(\mathbf{s};\mathbf{x})$ and the ground truth $u_0(\mathbf{s};\mathbf{x})$, and the emulation error is the error for the emulator $\hat{u}_N(\mathbf{s};\mathbf{x})$ given limited evaluations of the simulator $u_N(\mathbf{s};\mathbf{x})$.
To save the space, we investigate the numerical error and emulation error in Supplementary Materials S3 and S4, respectively, and then apply the error bound to a common PDE problem, \textit{elliptic equations}, in Supplementary Materials  S5.

\section{Numerical studies}\label{secnum}
Numerical studies are conducted in this section to examine the performance of the proposed method. Specifically, three real-world computer models comprised of PDEs are considered, which are solved via FEM using the quadratic shape functions as in \eqref{eq:quadraticshape}.

In these numerical studies, the hyperparameters in the priors \eqref{eq:gammadist} and \eqref{eq:musigmadist} can be quite generic without the need of estimation. Similar to \cite{yuan2008variational} and \cite{sun2010variational}, the  hyperparameter $\boldsymbol{\mu}_0$ in \eqref{eq:musigmadist} is set to the sample average of the node coordinates $\mathbf{S}_N$, and $\boldsymbol{\Sigma}_0$ are set to the sample inverse covariance of $\mathbf{S}_N$; the parameter $\kappa_0$ is the number of degrees of freedom under a Wishart distribution, which is set to the dimension of $\mathbf{s}_j$, that is, $\kappa_0=d$; the scale parameter $\mathbf{W}_0$ of the Wishart distribution is set to $\boldsymbol{\Sigma}_0/d$ such that the mean of $\boldsymbol{\Sigma}_k$ is the sample inverse covariance of $\mathbf{S}_N$. The concentration parameter $\alpha_0$ in \eqref{eq:gammadist} is set to 0.5.

For each individual GP, Mat\'ern correlation functions \eqref{matern} is considered. 
The smoothness parameter $\nu$ is set to $5/2$, which leads to a simplified form of \eqref{matern}:
\begin{equation*}\label{eq:maternkernel2.5}
    \Phi_{\bm{\theta}}(\mathbf{x}_i,\mathbf{x}_j)=\left(1+\sqrt{5}\|\mathbf{x}_i-\mathbf{x}_j\|_{\bm{\theta}}+\frac{5}{3}\|\mathbf{x}_i-\mathbf{x}_j\|_{\bm{\theta}}^2\right)\exp\left(-\sqrt{5}\|\mathbf{x}_i-\mathbf{x}_j\|_{\bm{\theta}}\right).
\end{equation*}
A small nugget parameter is added for numerical stability, which is set to $g\approx 1.5\times 10^{-8}$. The truncation level $K$ is set to 10. These numerical experiments were performed on a MacBook Pro laptop with Apple M1 Max of Chip and 32 GB of RAM.

Two performance measures are considered to examine the prediction performance. The first measure is the root mean squared error (RMSE) which is calculated as
\begin{equation}\label{eq:rmse}
{\rm RMSE}=\left(\frac{\sum^N_{j=1}\sum^{n_{\rm test}}_{i=1} (u_N(\mathbf{s}_j;\mathbf{x}^{\rm test}_i)-\hat{u}_N(\mathbf{s}_j;\mathbf{x}^{\rm test}_i))^2}{Nn_{\rm test}}\right)^{1/2},
\end{equation}
where $n_{\rm test}$ is the number of test input points, $u_N(\mathbf{s}_j;\mathbf{x}^{\rm test}_i)$ is the numerical solution of the $i$-th test input point, $\mathbf{x}^{\rm test}_i$, at the $j$-th node location, $\mathbf{s}_i$, and $\hat{u}_N(\mathbf{s}_j;\mathbf{x}^{\rm test}_i)$ is the posterior predictive mean as in \eqref{eq:mgGPemulator}. The second measure is the average continuous ranked probability score (CRPS) \citep{gneiting2007strictly}, which is a  performance measure for predictive distribution of a scalar observation. Since the predictive distribution is a mixture of normal distributions as in Section \ref{sec:prediction}, the CRPS can be computed by an analytical formula as in \cite{grimit2006continuous}. For both RMSE and CRPS, lower values indicate better prediction accuracy. 

To provide a comprehensive evaluation, we include the following methods for comparison: \texttt{uGP}, the independent univariate GPs sharing the same lengthscale hyperparameter ($\boldsymbol{\theta}_j$) but different scale hyperparameters ($\tau^2_j$) across the $N$ mesh nodes, which is similar to \cite{gu2016parallel}; \texttt{iGP}, the independent GPs sharing different hyperparameters ($\boldsymbol{\theta}_j$ and $\tau^2_j$) across the $N$ mesh nodes; \texttt{pcaGP}, which uses a functional principal component analysis (FPCA) with \textit{truncated} components \citep{wang2016functional}:
$$
u_{N}(\mathbf{s}_j;\mathbf{x}_i)\approx u_0(\mathbf{s}_j)+\sum^M_{l=1}\alpha_l(\mathbf{x}_i)\psi_l(\mathbf{s}_j),
$$
with the leading $M$ eigenfunctions $\{\psi_l(\mathbf{s})\}^M_{l=1}$, and the corresponding coefficients $\{\alpha_l(\mathbf{x}_i)\}$:
\begin{align*}\label{eq:klexpansion}
\psi_l(\mathbf{s}) &= \argmax_{\substack{\| \phi \|_2 = 1, \\ \langle \phi, \psi_l \rangle = 0, \forall l < j}} \sum^n_{i=1} \left\{ \int u_N(\mathbf{s};\mathbf{x}_i) \phi(\mathbf{s})  d\mathbf{s}\right\}^2, \nonumber\\
\alpha_l(\mathbf{x}_i) &= \int  u_N(\mathbf{s};\mathbf{x}_i) (\psi_l(\mathbf{s})-u_0(\mathbf{s})) \; d\mathbf{s},
\end{align*}
where $u_0(\mathbf{s})$ is the mean function, which can be estimated by $\sum^n_{i=1}u_N(\mathbf{s};\mathbf{x}_i)/n$. The number of components, $M$, is selected by finding the  leading $M$  eigenfunctions that explain  99\% of variance over all $n$ training cases. Then, \texttt{iGP} is applied to the $M$ coefficients $\{\alpha_l(\cdot)\}^M_{l=1}$. This approach is similar to \cite{dancik2008mlegp} and \cite{mak2017efficient}.

\subsection{Poisson's Equation}\label{sec:poisson}
\begin{figure}[t]
    \centering
    \includegraphics[width=0.8\textwidth]{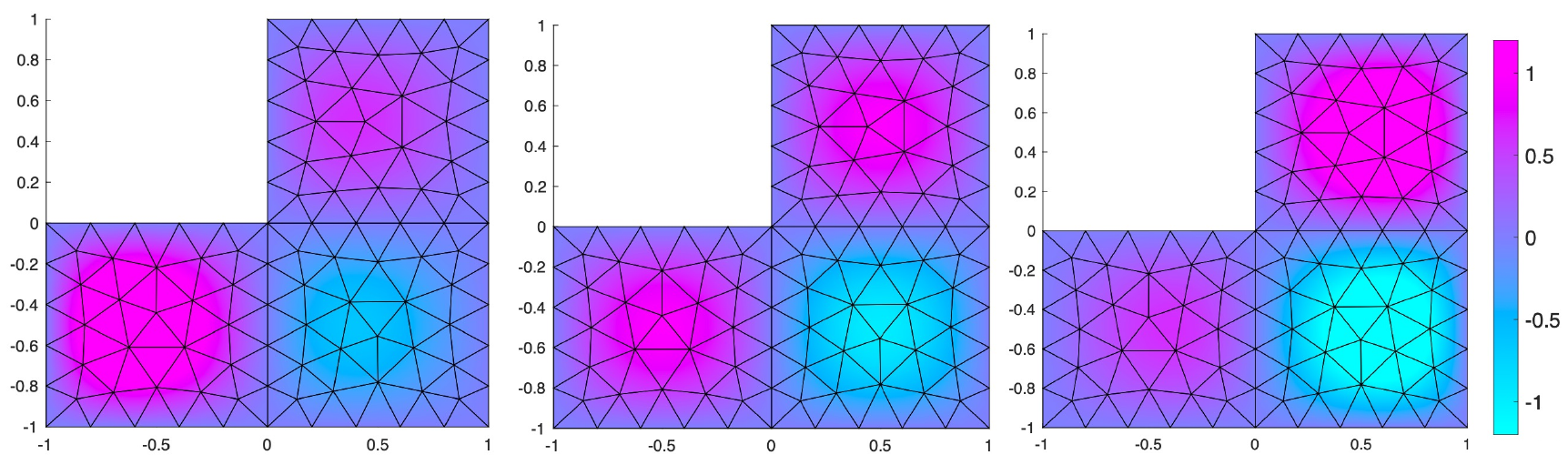}
    \caption{Illustration of geometry, mesh, and solutions via FEM for Poisson's equation with (left) $x=-1$, (middle) $x=0$, and (right) $x=1$.}
    \label{fig:PoissonDemo}
\end{figure}

In this subsection, we explore the performance  for emulating a PDE boundary value problem on an L-shaped membrane based on Poisson's equation, which has broad applicability in electrostatics and fluid mechanics  \citep{evans2010partial}. The model is represented as:
\begin{equation*}
    \Delta u=(x^2-2\pi^2)e^{xs_1}\sin(\pi s_1)\sin(\pi s_2)+2x\pi e^{xs_1}\cos(\pi s_1)\sin(\pi s_2),\quad \mathbf{s}=(s_1,s_2)\in \Omega,
\end{equation*}
where $x$ is the one-dimensional input variable with $x\in[-1,1]$, the operator $\Delta$  is defined by $\Delta=\frac{\partial^2}{\partial s_1^2}+\frac{\partial^2}{\partial s_2^2}$, $\Omega$ is an L-shaped membrane, and the Dirichlet boundary condition, $u=0$ on $\partial\Omega$, is considered. The geometry and mesh, along with the solutions through FEM when $x=-1,0,1$, are demonstrated in Figure \ref{fig:PoissonDemo}. Partial Differential Equation Toolbox of \cite{MATLAB:R2021b} is used to create the geometry and mesh to solve the equation. 

\begin{figure}[t]
    \centering
    \includegraphics[width=\textwidth]{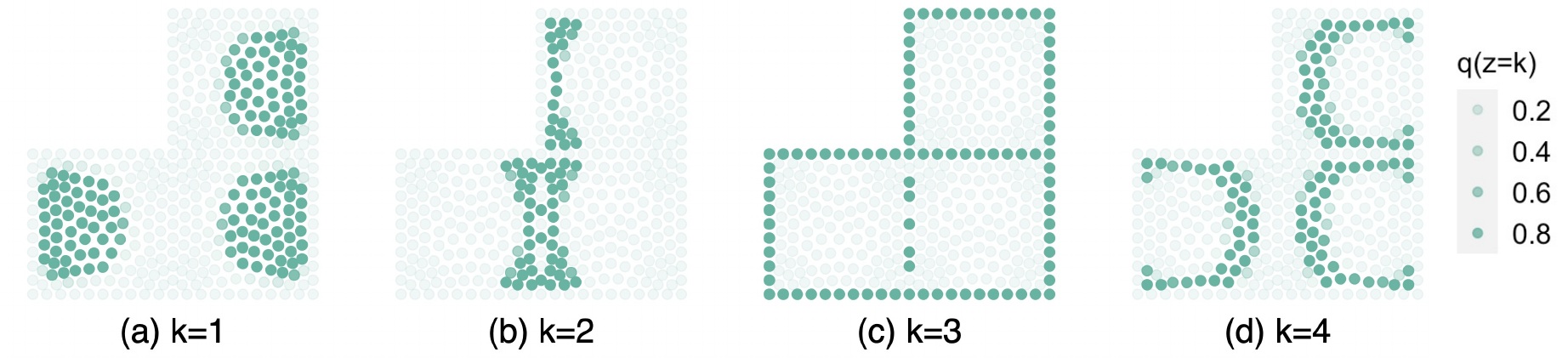}
    \caption{Variational distribution $q(z_j=k)$ in the Poisson's equation experiment. 
    }
    \label{fig:Poisson_qZ}
\end{figure}

\begin{figure}[t]
    \centering
    \includegraphics[width=0.55\textwidth]{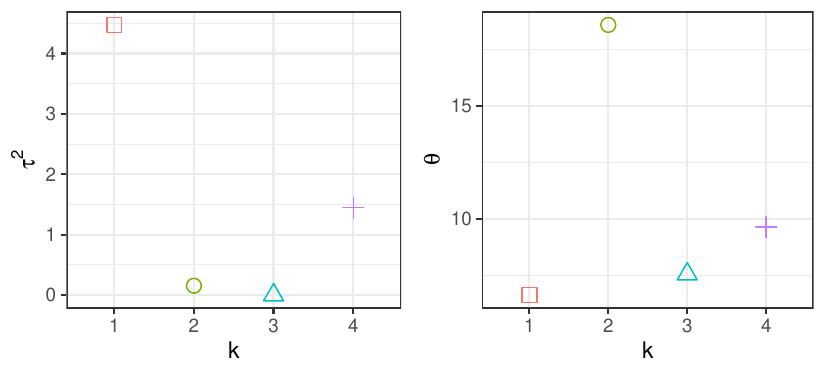}
    \caption{Hyperparameter estimates in the Poisson's equation experiment: (left) $\hat{\tau}_k$; (right)  $\hat{\theta}_k$.}
    \label{fig:PoissonVI}
\end{figure}

We conduct a computer experiment of size $n=5$, where the input locations are equally spaced in the input space $[-1,1]$, i.e., $x_i=0.4i-1.2$ for $i=1,\ldots,5$. The mesh size is set to 0.2, yielding $N=401$ mesh nodes as shown in Figure \ref{fig:PoissonDemo}. The proposed method is applied to this experimental data.  Figure \ref{fig:Poisson_qZ} illustrates the variational distribution,  $q(z_j=k)$, for $k=1,\ldots,4$, and Figure \ref{fig:PoissonVI} presents the corresponding  hyperparameter estimates, $\hat{\tau}_k^2$ and $\hat{\bm{\theta}}_k$. Note that here $k=5,6,\ldots,10$ are not shown in Figures \ref{fig:Poisson_qZ} and \ref{fig:PoissonVI} because $q(z_j=k)<0.001$ for all $j$, indicating that only four mixture components are only needed in this mixture model. This shows that, even though the proposed model considers an infinite mixture of GPs throughout a DP prior, the mixing proportions decrease exponentially so quickly that only a small number of components are used to model the data a priori. The fitting result reveals interesting scientific insights. First, the cluster $k=3$ has higher probability mass on the nodes at the boundaries and locations of $s_1=0$ and $s_2=0$ (Figure \ref{fig:Poisson_qZ}), and the corresponding hyperparamters of the GP are $\hat{\tau}^2_3\approx 0$ and $\hat{\theta}_3\approx 7.5$ (Figure  \ref{fig:PoissonVI}), which makes sense because these nodes are related to the solution of 0 based on the boundary condition. 
Second, the cluster $k=1$ features higher probability mass on the nodes in the regions where the magnitude of solutions is the highest and their shapes are similar. The cluster shares  high estimates $\hat{\tau}^2_1$ and $\hat{\theta}_1$, indicating the input-output relationship is smooth but the output values in these regions have relatively high variability across different input settings. More interestingly, the cluster $k=2$ covers a group of the nodes $s_1=0$ in the neighbourhood of the cluster $k=3$, while the cluster $k=4$ contains a group of nodes in the vicinity of the cluster $k=1$.
The example shows that, unlike the traditional reduced-basis methods like proper orthogonal decomposition (POD) \citep{lumley1967structure}, the clustering structures by the proposed method not only  provide a  useful insight for discovering the underlying physics, but also reveal their shared input-output relationship. 

\begin{figure}[h]
    \centering
    \includegraphics[width=0.9\textwidth]{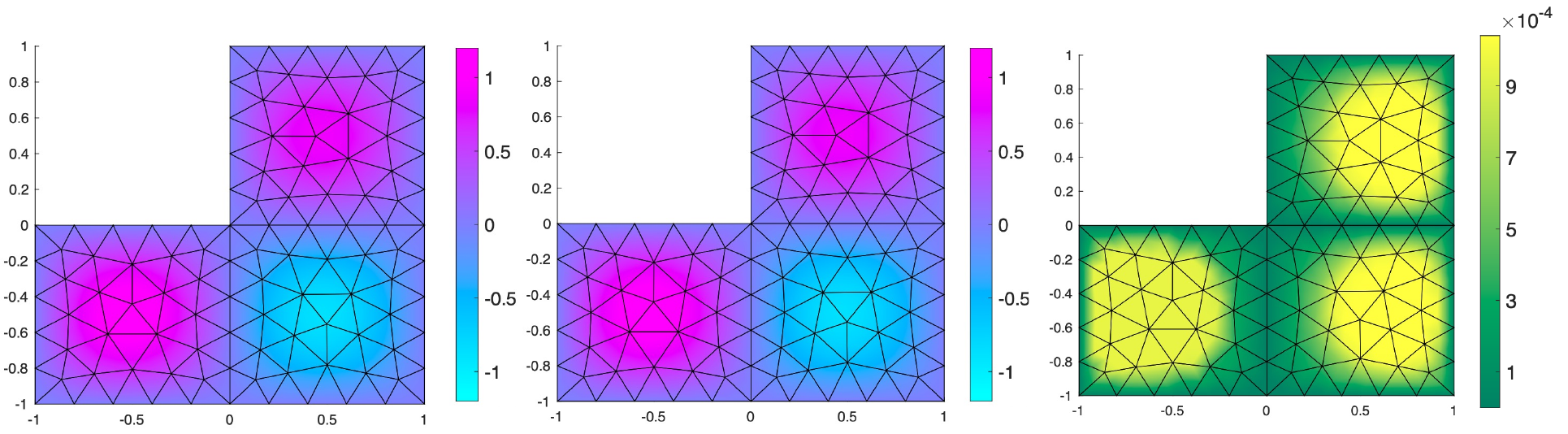}
    \caption{Validation performance of \textup{\texttt{mcGP}} prediction: (left) the real FEM solution when $x=-0.25$; (middle) the \textup{\texttt{mcGP}} posterior predictive mean; (right) the \textup{\texttt{mcGP}} posterior predictive standard deviation.}
    \label{fig:PoissonValidation}
\end{figure}

To illustrate the prediction performance, Figure \ref{fig:PoissonValidation} shows the FEM solution (left) at the test input point, $x^{\rm test}=-0.25$, the \texttt{mcGP} posterior predictive mean (middle), and the \texttt{mcGP} posterior predictive standard deviation (right). From visual comparison,  the \texttt{mcGP} posterior predictive mean accurately captures the spatial structure of the FEM solution. 
The prediction uncertainty can be quantified by the posterior predictive standard deviation, 
where the most uncertain predictions are located in the nodes of cluster $k=1$, which is expected provided that the variation of the outputs are most dynamic with large magnitude in this cluster. 

\begin{figure}[t]
    \centering
    \includegraphics[width=\textwidth]{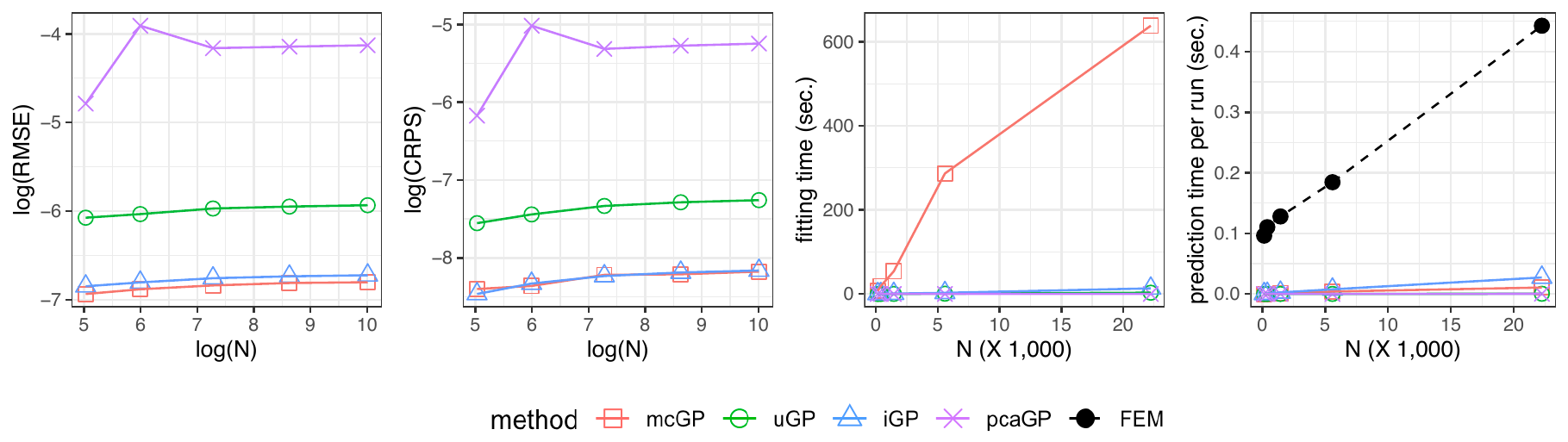}
    \caption{Performance comparison in terms of prediction accuracy and computational cost in the Poisson's equation experiment. The left two panels are the prediction performance in terms of RMSE and CRPS in logarithm, and the right two panels are the computational cost in model fitting and predictions.}
    \label{fig:poissoncomparison}
\end{figure}

We further examine the prediction and computation performance on 201 test input points equally spaced in $[-1,1]$ with various mesh size settings, which are 0.4, 0.2, 0.1, 0.05, and 0.025. The results are presented in Figure \ref{fig:poissoncomparison}. From the left two panels, it appears that \texttt{uGP} and \texttt{pcaGP} perform  worse than the other two in terms of prediction accuracy (both RMSE and CRPS). 
While there is no significant difference between \texttt{iGP} and \texttt{mcGP} in the predictive scores, it can be seen that \texttt{mcGP} generally yields lower RMSEs than other methods. The right two panels present the computational cost. It is of no surprise that the proposed method requires more fitting time due to the complications of a mixture model; however, as indicated in Section \ref{secParameter}, the third panel (from the left) shows that the computational cost for fitting \texttt{mcGP} is linear with respect to  $N$, which is reasonably tractable in practice. Besides, in the context of emulation, the computation for predictions is more of interest, for which it can be seen that the proposed method (and its competitors) can predict much faster than conducting a real FEM simulation. It is important to note that while in Figure 7, \texttt{iGP} demonstrates comparable prediction accuracy, in the problems presented in Sections \ref{sec:cylinder} and \ref{sec:blade}, it performs less accurately compared to \texttt{uGP} and \texttt{mcGP}. On the other hand, \texttt{mcGP} generally exhibits better prediction accuracy than \texttt{iGP} and \texttt{uGP}.


\subsection{Laminar flow past a cylinder}\label{sec:cylinder}

\begin{figure}[t!]
    \centering
    \includegraphics[width=0.9\textwidth]{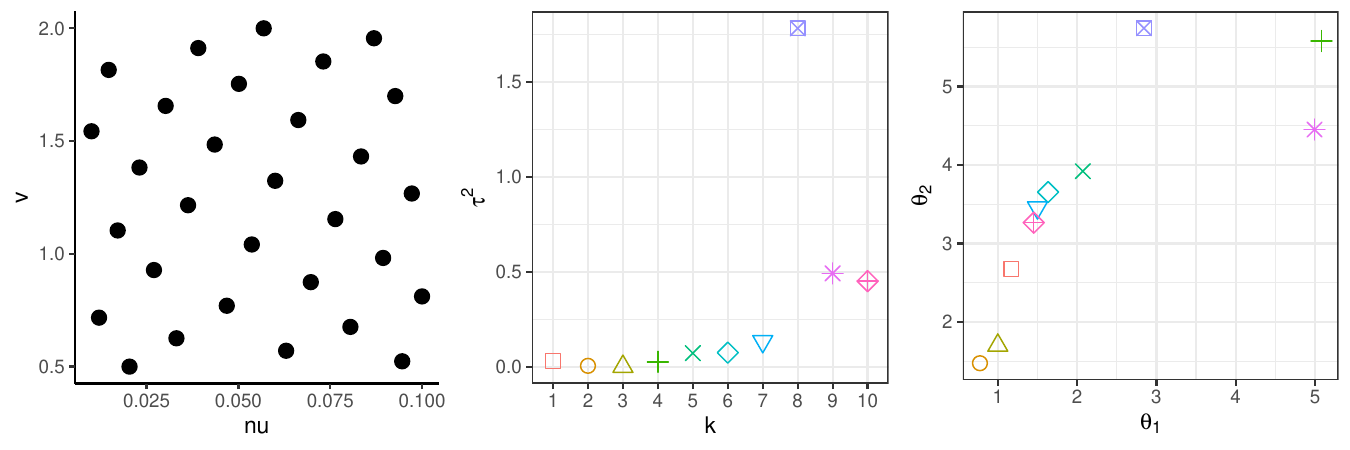}
    \caption{The left panel presents the design points in the laminar flow application, and the middle and right panels show the hyperparameter estimates of \textup{\texttt{mcGP}}: $\hat{\tau}_k$ (middle) and $\hat{\bm{\theta}}_k$ (right).}
    \label{fig:cylinderDOE}
\end{figure}

\begin{figure}[t!]
    \centering
    \includegraphics[width=\textwidth]{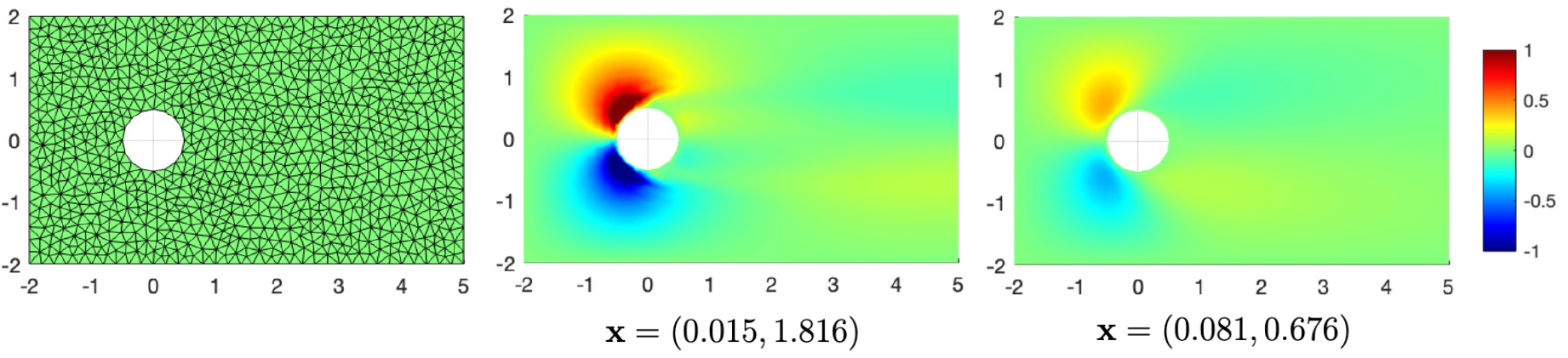}
    \caption{The left panel demonstrates the geometry and mesh in the laminar flow application, and the middle and right panels are the two (out of 30) training examples of the simulations, where the input settings are indicated by the values of $\mathbf{x}$. }
    \label{fig:cylinderdemo}
\end{figure}

In this subsection, we investigate a system of a two-dimensional flow past a circular cylinder, which is a classical and interesting problem in fluid mechanics \citep{rajani2009numerical,seo2012numerical}. The problem is described by the incompressible Navier-Stokes equations.
Two input variables are considered in this study, which are the kinematic viscosity of the fluid ($x_1$) and the freestream velocity in the $s_1$ direction ($x_2$), where the input space  is $\mathbf{x}=(x_1,x_2)\in[0.01,0.1]\times[0.5,2]$, which results in laminar flows with Reynolds number between 0.5 and 200. The objective is to predict the velocity component in the $s_2$ direction ($v$) using the \texttt{mcGP} method. The sample size is prescribed as $n=30$. The sample points are distributed uniformly in the input space following the maximum projection (MaxPro) design method \citep{joseph2015maximum}, as shown in the left panel of Figure \ref{fig:cylinderDOE}. The computer simulations at these points are performed using the FEM solver in the QuickerSim CFD Toolbox \citep{quickerSim}. Figure \ref{fig:cylinderdemo} shows the computational mesh and computer solutions of $v$ at two different input settings. The total number of mesh nodes is $N=3778$. 

\begin{figure}[h]
    \centering
    \includegraphics[width=\textwidth]{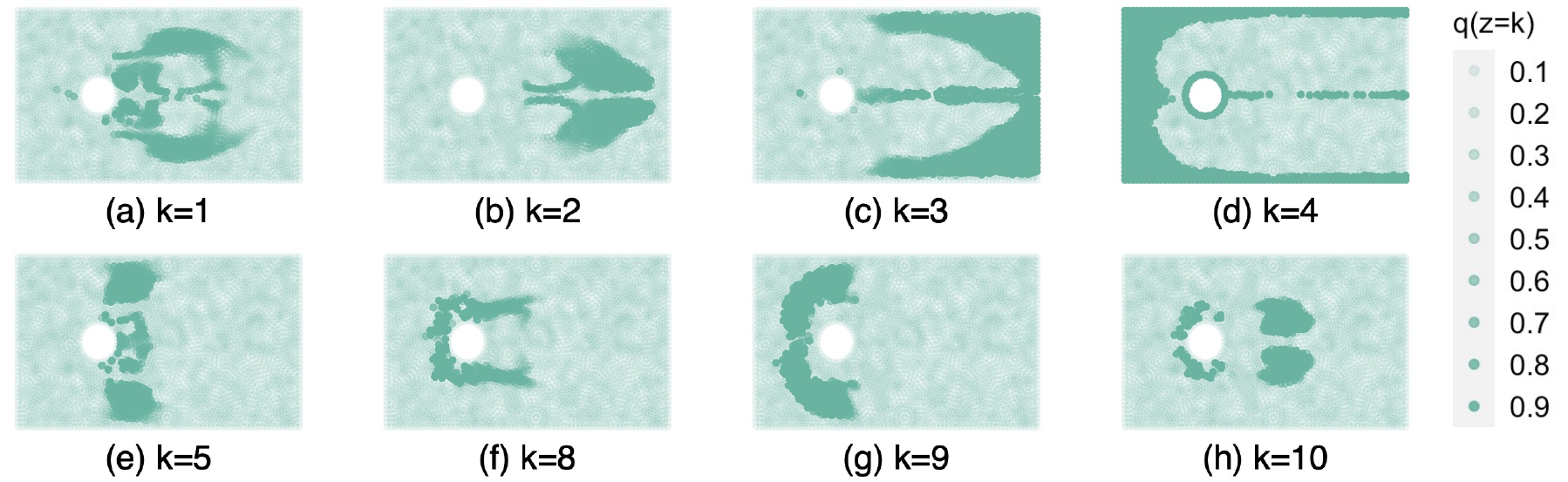}
    \caption{Variational distribution $q(z_j=k)$ in the laminar flow application.}
    \label{fig:cylinderqZ}
\end{figure}


The proposed \texttt{mcGP} method is applied to this experimental data.  Figure \ref{fig:cylinderqZ} illustrates the variational distribution, $q(z_j=k)$, for $j=1,\ldots,N$, and $k=1,\ldots,5,8,9,10$, and  the corresponding hyperparameter estimates are presented in the middle and right panels of Figure \ref{fig:cylinderDOE}  for each $k$. Note that the clusters $k=6$ and $k=7$ are not shown here because $q(z_j=k)<0.1$ for all $j$ at these clusters. The result shows that different clusters exhibit high probability mass on the nodes in different fluid regions. For example,  the cluster $k=4$ has higher probability mass on the nodes at the upstream and vertical boundaries with small $\hat{\tau}^2_4$ and large $\hat{\boldsymbol{\theta}}_4$, while the cluster $k=3$ imposes higher probability mass in the downstream boundary and associated neighbouring regions. These two clusters share a common feature of low variation in the vertical velocity, as manifested by Figure \ref{fig:cylinderdemo}. The components $k=8$ and $k=9$ puts higher probability mass in the frontal area of the cylinder with the highest output magnitude, and this comes with the large hyperparameter estimates $\hat{\tau}^2_8$ and $\hat{\tau}^2_9$. In addition, the clusters $k=1,2$, and $10$ have higher probability mass in the wake region. It turns out that the regions of high probability mass exerted by different clusters constitute the entire computational domain, implying the present clustering strategy is efficient. 

\begin{figure}[h]
    \centering
    \includegraphics[width=0.85\textwidth]{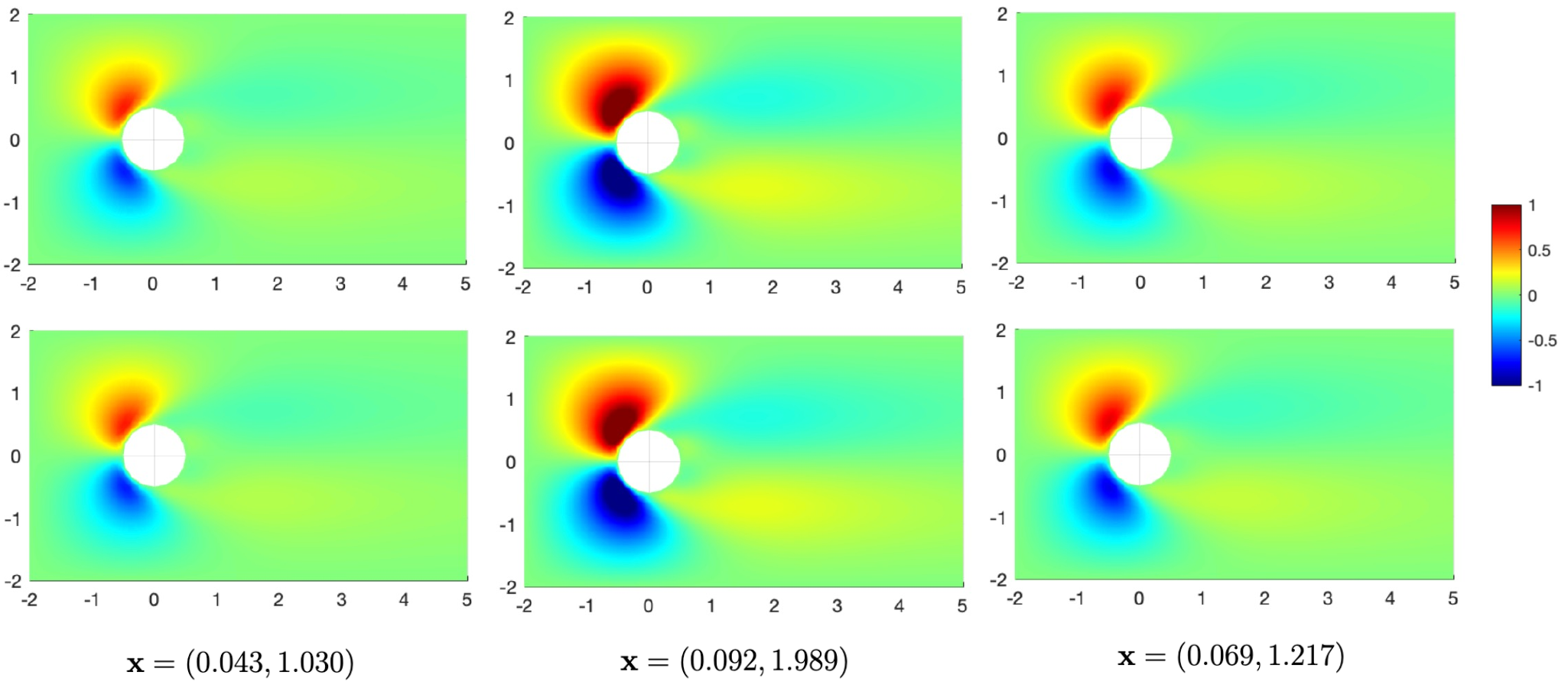}
    \caption{Validation performance of \textup{\texttt{mcGP}} prediction in the laminar flow application. The upper panels present three (out of 100) test FEM simulations (with the input settings indicated by the values of $\mathbf{x}$) and the bottom panels are the corresponding posterior predictive means of \textup{\texttt{mcGP}}.}
    \label{fig:cylindertest}
\end{figure}

The prediction and computation performance is examined on 100 uniformly random test input points from the input space. Three (out of 100) test FEM simulations along with the \texttt{mcGP} predictions are presented in Figure \ref{fig:cylindertest}. From visual inspection, it appears that the point-wise predictions of \texttt{mcGP} are fairly accurate at the three input points. All dynamic structures in the flow are well captured, including the large-magnitude region in the front of the cylinder and the wake region behind the cylinder. Table \ref{tbl:predictionperformance_cylinder} shows the results of the 100 test data in comparison of other GP models, indicating that the proposed \texttt{mcGP} outperforms the others in terms of prediction accuracy with reasonable computation time. Similar to the previous subsection, \texttt{pcaGP} performs the worst, which is not surprising because the approximation error of the dimension reduction approach can introduce additional bias to the predictions \citep{sung2022functional}. Unlike the previous subsection, \texttt{uGP} outperforms \texttt{iGP} in terms of both RMSE and CRPS. This again demonstrates that \texttt{mcGP} can serve as a middle ground between these two models.

\begin{table}[h]
    \centering
\begin{tabular}{ c|C{1.5cm}C{1.5cm}C{1.5cm}C{1.5cm} }
\toprule
{method}&\texttt{mcGP}&\texttt{uGP}&\texttt{iGP}&\texttt{pcaGP}\\
\midrule
RMSE ($\times 10^{-4}$) & \textbf{8.741} & 9.064 & 15.385 &24.880\\
CRPS ($\times 10^{-4}$) & \textbf{2.276} & 2.410 & 5.537 & 15.178\\ 
fitting time (sec.) & 98 & 10 & 10 & \textbf{0.13}\\
prediction time per run (msec.) & 3 & 0.1 & 19 & \textbf{0.06} \\
\bottomrule
\end{tabular}
\caption{Performance comparison in terms of prediction accuracy and computational cost in the laminar flow application, in which the better performances are boldfaced. Note that the test FEM simulations take, on average, 1068 milliseconds per run.}
\label{tbl:predictionperformance_cylinder}
\end{table}

\subsection{Thermal stress analysis of jet engine turbine blade}\label{sec:blade}
\begin{figure}[t!]
    \centering
    \includegraphics[width=0.9\textwidth]{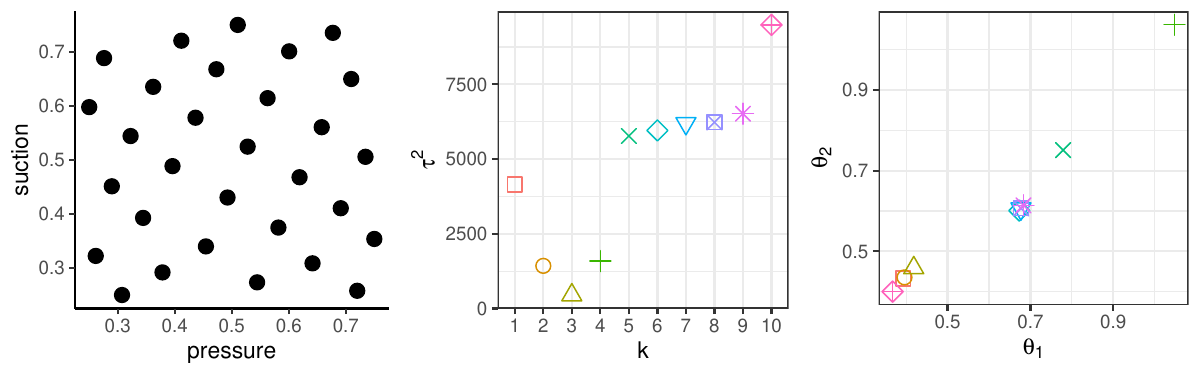}
    \caption{The left panel presents the design points in the turbine blade application, and the middle and right panels show the hyperparameter estimates of \textup{\texttt{mcGP}}: $\hat{\tau}_k$ (middle) and $\hat{\bm{\theta}}_k$ (right).}
    \label{fig:bladelDOE}
\end{figure}

\begin{figure}[t!]
    \centering
    \includegraphics[width=0.9\textwidth]{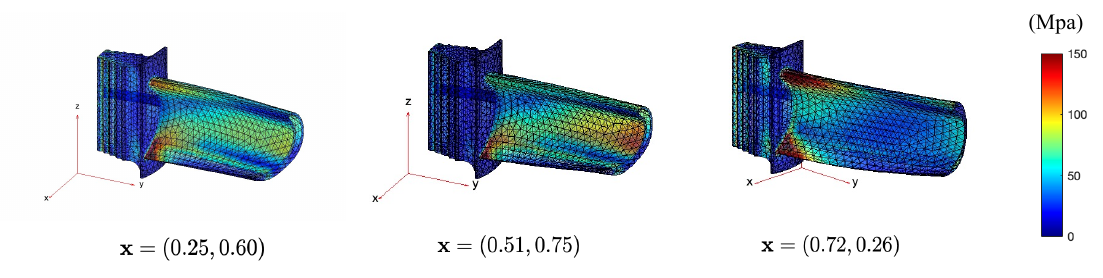}
    \caption{Three (out of 30) examples of the training FEM simulations in the turbine blade application, where the input settings are indicated by the values of $\mathbf{x}$.}
    \label{fig:bladeltraindemo}
\end{figure}

In this subsection, we investigate the performance of the proposed method on a thermal stress analysis application for a jet turbine engine blade in steady-state operating condition. More details can be found in  \cite{wright2006enhanced}, \cite{carter2005common} and \cite{sung2022stacking}. In this study, we aim to emulate the thermal stress and deformation of a turbine with the effect of the thermal stress and pressure of the surrounding gases on turbine blades. The problem is analyzed as a static structural model, which can be numerically solved using FEM. Two input variables are considered, which are the pressure load on the pressure ($x_1$) and suction ($x_2$) sides of the blade, both of which range from 0.25 to 0.75 MPa, i.e., $(x_1,x_2)\in[0.25,0.75]^2$. FEM simulations of sample size $n=30$ for the thermal stress are conducted via the Partial Differential Equation Toolbox \citep{MATLAB:R2021b}. The sample points  over the input space are allocated using the MaxPro design, which are presented in the left panel of Figure \ref{fig:bladelDOE}. The mesh size is set to 0.01, yielding $N=21158$ mesh nodes. Three  (out of 30) training examples are presented in Figure \ref{fig:bladeltraindemo}. They show that larger suction pressure ($x_2=0.60$ and $0.69$) tends to impose higher thermal stress on the central area of the blade airfoil and the blade leading and trailing edges near the platform. 


\begin{figure}[h]
    \centering
    \includegraphics[width=0.9\textwidth]{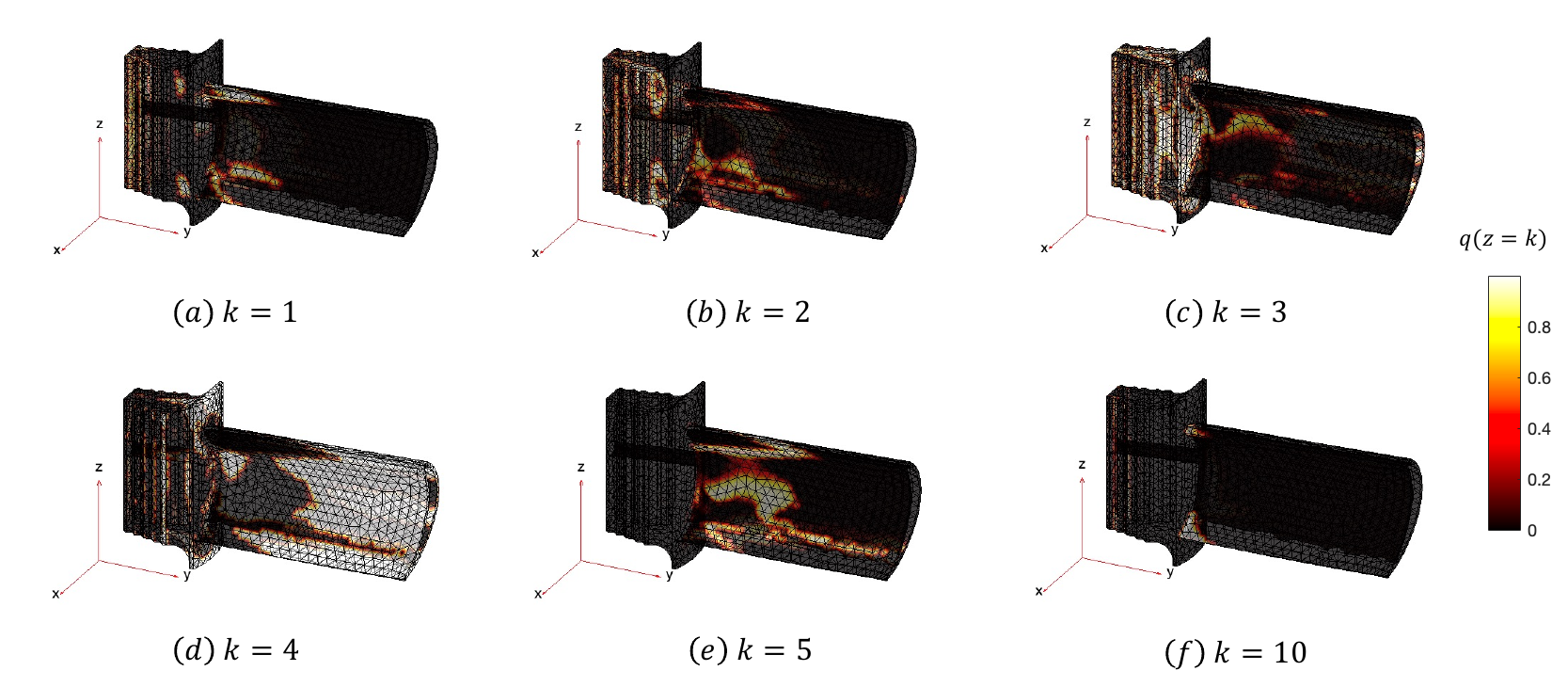}
    \caption{Variational distribution $q(z_j=k)$ in the turbine blade application.}
    \label{fig:bladeqZ}
\end{figure}

Figure \ref{fig:bladeqZ} illustrates the variational distribution of the selected components, $q(z_j=k)$, and the corresponding hyperparameter estimates, $\hat{\tau}_k^2$ and $\hat{\bm{\theta}}_k$, are shown in the middle and right panels of Figure \ref{fig:bladelDOE}. The clusters $k=6,7,8$ and $9$ are not shown here because $q(z_j=k)<0.3$ for all $j$ at these clusters. The distribution of the nodes in resulting clusters correspond to the distribution of thermal stress in a physically meaning manner. The cluster $k=4$ has higher probability mass on the top center of the blade airfoil, where the thermal load is relatively large with high variability ($\hat{\tau}^2_4\approx 1584$) across different input settings; on the other hand, the cluster $k=3$ put higher probability mass on the platform area, where the thermal load is relatively small with subtle variability (small $\hat{\tau}^2_3$). In addition, the cluster $k=1$ has higher probability mass on the leading and trailing edges of the blade next to the platform, which also tends to have a high variability ($\tau^2\approx 4147$) of the thermal stress.

\begin{figure}[h]
    \centering
    \includegraphics[width=0.9\textwidth]{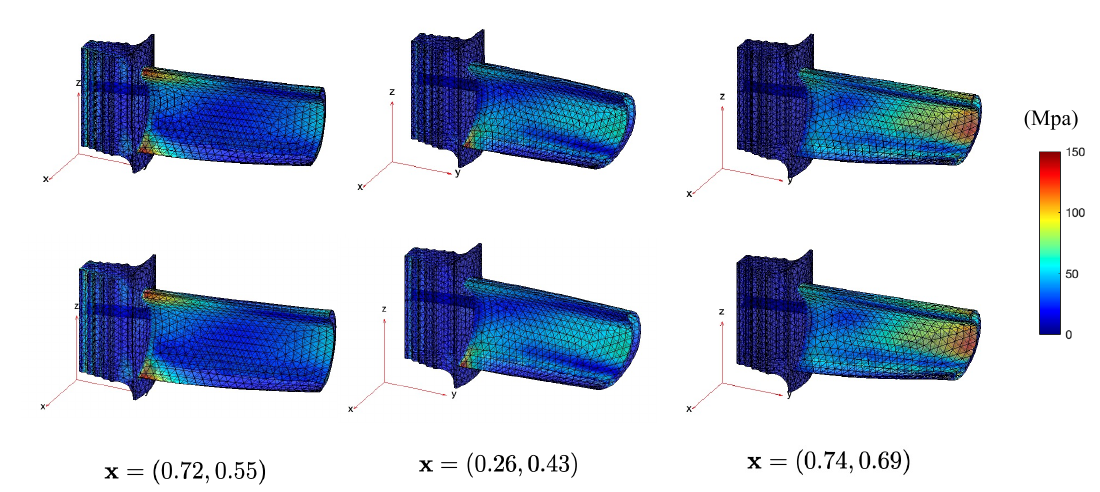}
    \caption{Validation performance of \textup{\texttt{mcGP}} prediction in the turbine blade application. The upper panels present three (out of 100) test FEM simulations (with the input settings indicated by the values of $\mathbf{x}$) and the bottom panels are the corresponding posterior predictive means of \textup{\texttt{mcGP}}.}
    \label{fig:bladeltestdemo}
\end{figure}

Similar to Section \ref{sec:cylinder}, 100 test FEM simulations are conducted at uniformly random test input locations  to examine the prediction and computation performance. Three (out of 100) test examples are demonstrated in Figure \ref{fig:bladeltestdemo}, which shows that, from visual inspection, the predicted thermal stress (upper panels) closely mimics the simulated thermal stress (lower panels). In comparison with the competitors, the results are presented in Table \ref{tbl:predictionperformance_blade}, which again shows  that the proposed method can outperform others in terms of prediction accuracy with reasonable computation time.

\begin{table}[h]
    \centering
\begin{tabular}{ c|C{1.5cm}C{1.5cm}C{1.5cm}C{1.5cm} }
\toprule
{method}&\texttt{mcGP}&\texttt{uGP}&\texttt{iGP}&\texttt{pcaGP}\\
\midrule
RMSE ($\times 10^{-1}$)& \textbf{9.800} & 9.823 & 11.073 & 13.986\\
CRPS ($\times 10^{-1}$)& \textbf{3.204} & 3.236 & 3.617 & 6.753\\ 
fitting time (sec.) & 313 &134 & 71& \textbf{0.19}\\
prediction time per run (msec.) & 19 & 0.7 & 97 & \textbf{0.3} \\
\bottomrule
\end{tabular}
\caption{Performance comparison in terms of prediction accuracy and computational cost in the turbine blade application, in which the better performances are boldfaced. Note that the test FEM simulations take, on average, 3819 milliseconds per run.}
\label{tbl:predictionperformance_blade}
\end{table}

\section{Concluding remarks}\label{secconclu}
PDE simulations based on mesh-based numerical methods have become essential in various applications, ranging from engineering to health care. In this paper, we propose a new surrogate model for expensive PDE boundary value problems that simultaneously emulates the PDE solutions over a spatial domain. An important innovation of this work lies in its incorporation of mesh node locations into a statistical model, forming a mixture model that compromises the bias-variance trade-off in the context of many outputs. Furthermore, we develop a rigorous theoretical error analysis for the proposed emulator, which provides an important insight about its uncertainty quantification. Three real examples show that the method not only has advantages in prediction accuracy, but also enables discovery of interesting physics by interpreting the mixture clusters.

The proposed method shows several avenues for future research. First, in addition to the fine mesh in a spatial domain, the proposed method can be modified for simulations having fine grids over both the spatial and temporal domains. This can be naturally done by incorporating the time information into the latent model  \eqref{eq:nodedist} to segment the temporal domain, and it is conceivable that the resulting clustering structures can reveal interesting \textit{dynamic} features. Second, although the method developed herein assumes that the mesh specifications are identical across different input settings, the proposed method can be modified to tackle different mesh specifications. This can be done by utilizing the idea of \textit{common grid} by \cite{mak2017efficient}. 
Specifically, we recommend first determining a common grid, denoted as $\bar{\mathbf{S}}=(\bar{\mathbf{s}}_1,\ldots,\bar{\mathbf{s}}_N)^T$, which serves as a reference for predictions. By evaluating training outputs $u_N(\bar{\mathbf{s}}_j;\mathbf{x}_i)$ for each $i=1,\ldots,n$ and $j=1,\ldots,N$ via \eqref{PDEnum}, we can subsequently model $\{u_N(\bar{\mathbf{s}}_j;\mathbf{x})\}_{j=1,\ldots,N}$ as an \texttt{mcGP}, as introduced in Section \ref{secMethod}. This strategy enables  predictions at untried points $\mathbf{x}\in\chi$ on the common grid $\bar{\mathbf{S}}$. Lastly, it is worthwhile investigating the incorporation of the boundary conditions into the proposed model as in  \cite{tan2018gaussian,tan2018gaussianss},  to further improve the prediction accuracy. We leave these to our future work.

\vspace{0.5cm}
\noindent\textbf{Supplemental Materials}
Additional supporting materials can be found in Supplemental Materials, including the detailed derivations for Algorithm \ref{alg:variationalEM}, the derivation of \eqref{eq:postvar}, the detailed error analysis in Section  \ref{secError}, and the \textsf{R} code for reproducing the results in Section
\ref{secnum}.

\bibliography{ref}

\end{document}